\documentclass[preprint,aps,amsmath,nofootinbib,tightenlines]{revtex4}
\usepackage{graphicx}
\usepackage{bm}
\usepackage{epsfig}
\usepackage[dvips]{hyperref}

\newif\ifpdf
\ifx\pdfoutput\undefined
\pdffalse 
\else
\pdfoutput=1 
\pdftrue
\fi

\def\nslash{n\!\!\!\slash}
\def\bnslash{\bar n\!\!\!\slash}

\def\OMIT#1{}

\newcommand{\nn}{\nonumber} 

\newcommand{\bn}{{\bar n}}
\newcommand{\bea}{\begin{eqnarray}}
\newcommand{\eea}{\end{eqnarray}}

\newcommand{\mcdot}{\!\cdot\!}

\def\babar{\mbox{\sl B\hspace{-0.4em} {\small\sl A}\hspace{-0.37em} \sl
B\hspace{-0.4em} {\small\sl A\hspace{-0.02em}R}\hspace{0.3em}}}

\begin{document}
\setlength\baselineskip{15pt}

\ifpdf
\DeclareGraphicsExtensions{.pdf, .jpg}
\else
\DeclareGraphicsExtensions{.eps, .jpg}
\fi


\preprint{ \vbox{ \hbox{arXiv:0802.0873} \hbox{CALT-68-2664} }}

\title{\phantom{x}
\vspace{0.5cm}
Subleading Shape-Function Effects
and the Extraction of $|V_{ub}|$
\vspace{0.6cm}
}

\author{Keith S. M. Lee}
\affiliation{California Institute of 
Technology,
Pasadena, CA 91125\footnote{Electronic address: ksml@caltech.edu}
\vspace{0.5cm}}
\vspace{1cm}

\begin{abstract}
\vspace{0.3cm}
We derive a class of formulae relating moments of $B\to X_u\ell\bar\nu$ 
to $B\to X_s\gamma$ in the shape function region, where 
$m_X^2 \sim m_b \Lambda_{\rm QCD}$.
We also derive an analogous class of formulae involving the
decay $B\to X_s\ell^+\ell^-$.
These results incorporate $\Lambda_{\rm QCD}/m_b$ power corrections,
but are independent of leading and subleading hadronic shape functions.
Consequently, they enable one to determine $|V_{ub}|/|V_{tb}V_{ts}^*|$ 
to subleading order in a model-independent way.
\end{abstract}

\maketitle

\section{Introduction} \label{intro}

The study of decays of the $B$ meson allows us to probe QCD and 
flavour physics. The program's goals include, on the one hand,
precision measurements of Standard Model parameters and, on the
other hand, searches for new physics.
Short-distance physics is encoded in Wilson coefficients of local
operators. By comparing measurements of these coefficients with
theoretical predictions, signals of new physics may be found.
High sensitivity to new physics is provided by the so-called
rare decays, namely those channels involving flavour-changing
neutral currents, since they do not occur at tree level in the
Standard Model. Measurements of the inclusive rare process
$B\to X_s\gamma$~\cite{cleo95,aleph98,belle01,cleo01,babar02} 
have provided significant constraints on 
extensions to the Standard Model. The more complicated decay
$B\to X_s\ell^+\ell^-$ is complementary to $B\to X_s\gamma$,
as its effective Hamiltonian includes two extra operators. 
Moreover, additional observables are available, such as the
$q^2$ spectrum and the forward-backward asymmetry, which have
been the focus of much work. Recently, it was noted that an
angular decomposition provides a third observable sensitive
to a different combination of Wilson coefficients~\cite{Lee:2006gs}. 
Belle and \babar have already made initial measurements of 
$B\to X_s\ell^+\ell^-$~\cite{belle03,babar03}.

Precision measurements also provide determinations of elements
of the CKM matrix or, equivalently, the angles and sides of the 
unitarity triangle. By overconstraining these, the flavour structure
of the Standard Model is subjected to rigorous examination. For
the decay $B\to X_c\ell\bar\nu$, experimental and theoretical
uncertainties are under control, and consequently $|V_{cb}|$
is one of the best-determined elements of the CKM matrix. From
$B\to X_u\ell\bar\nu$, we can also determine 
$|V_{ub}|$~\cite{Aubert:2005mg,Golubev:2007cs,Bizjak:2005hn,Gibbons:2004dg}.

However, inclusive $B$ decays often require a trade-off between
theoretical and experimental difficulty: if phase-space cuts are
necessary experimentally, then the spectra will be less inclusive
and the corresponding theory more complicated. In this respect,
$B\to X_c\ell\bar\nu$ and $B\to X_u\ell\bar\nu$ are markedly 
different. The former is sufficiently inclusive to enable the
use of a local operator product expansion (OPE)~\cite{Shifman}, 
in which non-perturbative corrections appear as an expansion in 
inverse powers of $m_b$. This formalism has been calculated to order
$1/m_b^3$~\cite{Gremm:1996df} (and recently to order 
$1/m_b^4$~\cite{Dassinger:2006md}), with the relevant 
non-perturbative matrix elements defined via the Heavy Quark Effective 
Theory (HQET)~\cite{Grinstein:1990mj,Eichten:1989zv,Georgi:1990um}. 
In contrast, in  $B\to X_u\ell\bar\nu$ experimental cuts (e.g. cuts on 
$E_\ell$ or $m_X^2$) are required in order to eliminate the dominant
$b\to c$ background. In many cases, we are restricted to a region
in which $m_X^2 \sim m_b \Lambda_{\rm QCD}$ and the local OPE
breaks down. In this so-called endpoint or shape function 
region~\cite{Neubert:1993ch}, the set of outgoing hadronic states 
becomes jet-like and the relevant degrees of freedom are collinear and 
ultrasoft modes.  The Soft-Collinear Effective Theory 
(SCET)~\cite{bfl01,bfps01,bs1,bps02} is then a powerful 
theoretical method.

Similarly, $B\to X_s\gamma$ measurements employ a cut on the photon energy. 
In Refs.~\cite{Lee:2005pk,Lee:2005pw} 
it was shown that the shape function region is also relevant for 
$B\to X_s\ell^+\ell^-$.  Here, cuts are made in the dileptonic mass 
spectrum to remove the largest $c\bar c$ resonances, namely the 
$J/\Psi$ and $\Psi'$.
These leave two perturbative windows, the low-$q^2$ and high-$q^2$
regions. At low $q^2$, where the rate is higher, an additional cut
is needed: a hadronic invariant-mass cut is imposed in order to
eliminate the background $b\to c(\to s \ell^+ \nu) \ell^-\bar\nu$.

At leading order (LO) in $\Lambda_{\rm QCD}/m_b$, decay rates now depend
upon a non-perturbative, and hence analytically incalculable,
shape function. However, this function is process-independent and
appears in both $B\to X_u\ell\bar\nu$ and $B\to X_s\gamma$, for example. 
One can thus measure the leading-order shape function from the photon energy
spectrum of $B\to X_s\gamma$ and use the result in the $B\to X_u\ell\bar\nu$
spectrum, or, more directly, express the semileptonic rate in terms of the 
radiative rate instead of the shape function%
~\cite{Leibovich:1999xf,Leibovich:2000ey,Lange:2005qn,Lange}.
In this way, model dependence can be avoided in the determination of
$|V_{ub}|$.

At subleading order, the situation is far more complicated, with 
several universal shape functions occurring in different combinations%
~\cite{blm1,blm2,klis04,Bosch,Beneke,Trott}.
In this paper, we construct combinations of shape-function-dependent
decay rates that are protected from non-perturbative effects to
second order in the power expansion. Through this procedure, we obtain
formulae for $|V_{ub}|/|V_{tb} V_{ts}^*|$ that are free from the hadronic 
uncertainties  arising from the leading and subleading shape functions.
This method uses moments of the fully differential decay spectra
of $B\to X_u\ell\bar\nu$ and $B\to X_s\gamma$ 
(and, optionally, $B\to X_s\ell^+\ell^-$).

The rest of this paper is organized as follows. In Sec.~2, together
with Appendices~A and B, we present the basic formalism needed for our work. 
This includes power corrections for the triply differential decay spectra
of the semileptonic processes and the photon energy spectrum of 
$B\to X_s\gamma$. In Sec.~3, we derive and discuss our results, 
eliminating shape functions from expressions for $|V_{ub}|$ at 
next-to-leading order (NLO). We conclude in Sec.~4.

\section{formalism} \label{formalism}

In this section, we briefly review the formalism and results from 
Refs.~\cite{klis04,Lee:2005pk,klft07} that we shall use in this paper 
(see these references for further details).

The inclusive decay rate for $\bar B\to X_u \ell \bar\nu$ 
($\bar B\to X_s \gamma$) is proportional to $W_{\mu\nu} L^{\mu\nu}$, where
$L^{\mu\nu}$ is the leptonic (photonic) tensor and  $W_{\mu\nu}$ is the
hadronic tensor, which can be written as
\begin{eqnarray} \label{defnW}
  W_{\mu\nu} &=& \frac{1}{2m_B} \sum_X (2\pi)^3 \delta^4(p_B-q-p_X)
  \langle \bar B | J_\mu^\dagger | X \rangle \langle X | J_\nu | \bar B \rangle
  \nn\\
  &=& - g_{\mu\nu} W_1 + v_\mu v_\nu W_2 + i\epsilon_{\mu\nu\alpha\beta}
  v^\alpha q^\beta W_3 + q_\mu q_\nu W_4 + (v_\mu q_\nu + v_\nu q_\mu) W_5
  \,.
\end{eqnarray}
Here, $v^\mu$ is the velocity of the $B$ meson and $q^\mu$ is the 
$\ell\bar\nu$ ($\gamma$) momentum. 
We use the hadronic current $J$ 
(e.g.\ $J_\mu^u= \bar u\, \gamma_\mu P_L b$ for $B\to X_u\ell\bar\nu$) 
and relativistic normalization for the $|\bar B\rangle$ states.  
Similarly, the  inclusive decay rate for 
$\bar B\to X_s \ell^+ (p^+)\ell^- (p^-)$
is proportional to $(W^L_{\mu\nu}L_L^{\mu\nu} + W^R_{\mu\nu}L_R^{\mu\nu})$,
where 
$L_{L(R)}^{\mu\nu} =  2 \left[ p_+^\mu \, p_-^\nu + p_-^\mu \, p_+^\nu
   - g^{\mu\nu}\, p_+ \!\cdot \! p_- \mp i \epsilon^{\mu\nu\alpha\beta} \,
                      p_{+\alpha} \, p_{-\beta} \right]$
and $W^{L(R)}_{\mu\nu}$ can be defined analogously to Eq.~(\ref{defnW}),
in terms of a current $J^{L(R)}$ \cite{ahhm97}.

Contracting  $L^{\mu\nu}$ with $W^{\mu\nu}$ and 
neglecting the mass of the leptons give the differential decay rates
\begin{eqnarray} \label{dGamma}
  \frac{d\Gamma^s}{dE_\gamma}
 &=& \Gamma_0^s\: \frac{8 E_\gamma}{m_B^3} (4 W_1^s -W_2^s -2 E_\gamma W_5^s)  
    \,, \\[5pt]
 \frac{d^3\Gamma^u}{ dE_\ell dq^2 dE_\nu}  
    &=& \Gamma_0^u\: \frac{96}{m_B^5} \Big[ q^2 W_1^u + 
    (2 E_\ell E_\nu -q^2/2) W_2^u  + q^2 (E_\ell-E_\nu) W_3^u \Big]
    \theta(4 E_\ell E_\nu-q^2) \,, \nn\\[5pt] 
%
 \frac{d^3\Gamma^{\ell\ell}}{dq^2 dE_{+} dE_{-}}
    &=& \Gamma_0^{\ell\ell}\: \frac{96}{m_B^5} \Big[ q^2 W_1^{\ell\ell} +
   (2 E_{-} E_{+} -q^2/2) W_2^{\ell\ell}
    + q^2 (E_{-}-E_{+}) W_3^{\ell\ell} \Big]
 \theta(4 E_- E_+ -q^2)
    \,, \nn
\end{eqnarray}
for $B\to X_s\gamma$, $B\to X_u\ell\bar\nu$ and $B\to X_s\ell^+\ell^-$,
respectively,
where $W_{1(2)}^{\ell\ell}=W_{1(2)}^L+W_{1(2)}^R$, 
$W_3^{\ell\ell}=W_3^L-W_3^R$,
$W_i=W_i(q^2,v\mcdot q)$ and the normalization factors are
\begin{eqnarray}
  \Gamma_0^s & = & \frac{G_F^2\, m_B^3}{32\pi^4}\,
  |V_{tb} V_{ts}^*|^2\, \alpha_{\rm em}\, [\overline m_b(m_b)]^2
  |C_7^{\rm eff(0)}(m_b)|^2 \,,\qquad
 \Gamma_0^u = \frac{G_F^2\, m_B^5}{192\pi^3}\,
  |V_{ub}|^2\,, \nn \\
 \Gamma_0^{\ell\ell} & = & \frac{G_F^2\, m_B^5}{192\pi^3}\,
  \frac{\alpha_{\rm em}^2}{16\pi^2}\, |V_{tb} V_{ts}^*|^2\,.
%
\end{eqnarray}
In SCET, it is natural to use light-cone coordinates, where we
introduce vectors $n$ and $\bn$ such that $n^2=\bn^2=0\;$ and
$n\mcdot\bn = 2$. A four-vector then has components 
$(p^+, p^-, p_\perp) = (n\cdot p, \bn\cdot p, p_\perp^\mu)$.
In the region of interest, the set of hadronic states $X$ is jet-like, 
i.e.\ $p_X^+ \ll p_X^-$.
For convenience we define the dimensionless variables
\begin{eqnarray} \label{ybaru}
  x_H^\gamma = \frac{2 E_{\gamma}}{m_B} \,, \qquad
  x_H = \frac{2 E_{\ell}}{m_B} \,, \qquad
  \overline y_H = \frac{\bn\mcdot p_X}{m_B} \,,\qquad 
  u_H = \frac{n\mcdot p_X}{m_B} \,.
\end{eqnarray}
In terms of these variables, the decay rates are
\begin{eqnarray} \label{dGamma3u}
\frac{d\Gamma^s}{dx_H^\gamma}
    &=& \Gamma_0^s\: \frac{2 x_H^\gamma}{m_B}
     \Big\{ 4 W_1^s -W_2^s -m_B\, x_H^\gamma W_5^s \Big\}
    \,, \\
 \frac{1}{\Gamma_0^u} \frac{d^3\Gamma^u}{ dx_H\, d\overline y_H\, du_H}  
    &=& \: {24m_B} (\overline y_H\!-\! u_H)
   \Big\{(1\!-\! u_H)(1\!-\! \overline y_H) W_1^u + 
    \frac12 (1\!-\!x_H\!-\!u_H)(x_H\!+\!\overline y_H\!-\!1) W_2^u \nn\\
 && \hspace{1cm}
  + \frac{m_B}{2}\,(1\!-\! u_H)(1\!-\! \overline y_H)
     \big(2x_H\!+\! u_H \!+\!\overline y_H\!-\! 2\big) W_3^u \Big\}
    \,, \nn \\
 \frac{1}{\Gamma_0^{\ell\ell}} 
 \frac{d^3\Gamma^{\ell\ell}}{ dx_H\, d\overline y_H\, du_H}  
    &=& \: {24m_B} (\overline y_H\!-\! u_H)
   \Big\{(1\!-\! u_H)(1\!-\! \overline y_H) W_1^{\ell\ell} + 
    \frac12 (1\!-\!x_H\!-\!u_H)(x_H\!+\!\overline y_H\!-\!1) 
    W_2^{\ell\ell} \nn\\
 && \hspace{1cm}
  + \frac{m_B}{2}\,(1\!-\! u_H)(1\!-\! \overline y_H)
     \big(2x_H\!+\! u_H \!+\!\overline y_H\!-\! 2\big) 
    W_3^{\ell\ell} \Big\}
    \,, \nn 
\end{eqnarray}
where $W_i=W_i(u_H,\overline y_H)$. 
The full phase-space limits are given in Table~II of Ref.~\cite{klis04}.

The optical theorem relates the $W_i$ to forward-scattering amplitudes,
which can be calculated by taking time-ordered products of currents.
An important part of the analysis is the separation of short- and 
long-distance contributions.  The results, known as factorization 
theorems, may be written schematically in the form
\begin{align}
 d \Gamma &= H \times {\cal J} \otimes f \,, \nn
\end{align}
where $\otimes$ denotes a convolution. The hard ($H$) and jet 
(${\cal J}$) functions encode perturbative corrections that appear at
two different scales, $\mu_b\sim m_b$ and 
$\mu_i \sim \sqrt{m_b\Lambda_{\rm QCD}}$ respectively, 
whereas the shape function ($f$) represents non-perturbative physics.

SCET involves a power expansion in the small parameter 
$\lambda = \sqrt{\Lambda_{\rm QCD}/m_b}$. At leading order in $\lambda$,
rates depend on one shape function, which we denote by $f^{(0)}$:
\begin{align} 
W_i^{(0)} &=
  h_i(p_X^-,m_b,\mu) \:
  \int_{0}^{p_X^+} \! \! dk^+\:
   {\cal J}^{(0)}(p_X^-\, k^+,\mu )\:
   f^{(0)}(k^+  \!+\!\overline\Lambda\!-\! p_X^+,\mu) \,,
\end{align}
where 
$\overline\Lambda = m_B - m_b + (\lambda_1+3\lambda_2)/(2m_b) + \ldots$. 
The first subleading shape functions occur at order $\lambda^2$ and we
denote these by $f_{0-2}^{(2)}$, $f_{3,4}^{(4)}$ and $f_{5,6}^{(6)}$. 
These are common to the three decays, but appear in different combinations,
and are convoluted with jet functions ${\cal J}^{(0)}$, ${\cal J}^{(-2)}$
and ${\cal J}^{(-4)}$, respectively, as shown in Eq.~(\ref{Wfact00c}).
Note that we also have $u_H/\bar y_H \sim \lambda^2$
in the shape function region.

The shape functions are given by $B$-meson matrix elements of non-local 
ultrasoft operators. 
The definitions used here follow Ref.~\cite{klis04} and are included
in Appendix~\ref{app:sf}.
At tree level, the jet functions are
\begin{eqnarray}
 && {\cal J}^{(0)}(k^+) = \delta(k^+) \,,\qquad
  {\cal J}^{(-2)}(k_j^+)=\frac{\delta(k_1^+)-\delta(k_2^+)}{k_2^+ - k_1^+}
   \,,\\
 && {\cal J}^{(-4)}(k_j^+) =
  {4\pi\alpha_s(\mu_i)}
  \left[\frac{\delta(k_1^+)}{(k_2^+)(k_3^+)}
      + \frac{\delta(k_2^+)}{(k_1^+)(k_3^+)}
      + \frac{\delta(k_3^+)}{(k_1^+)(k_2^+)}
      - \pi^2 \delta(k_1^+)\delta(k_2^+)\delta(k_3^+)
      \right]  .\nn
\end{eqnarray}
At one-loop order, we have
\begin{eqnarray} \label{J01loop}
  {\cal J}^{(0)}(\omega,k^+,\mu) &=&
   \bigg\{ \delta(k^+) \bigg[
   1 + \frac{\alpha_s(\mu)C_F}{4\pi} \Big( 2 \ln^2\frac{\omega p_X^+}{\mu^2} -
   3 \ln \frac{\omega p_X^+}{\mu^2} + 7-\pi^2 \Big) \bigg] \\
 && +\frac{\alpha_s(\mu)C_F}{4\pi}
  \bigg[ \Big(\frac{4\ln (k^+/p_X^+)}{k^+}\Big)_+ +
     \Big(4 \ln\frac{\omega p_X^+}{\mu^2}-3 \Big)\: \frac{1}{(k^+)_+} \bigg]
   \bigg\}
   \theta(p_X^+\!-\!k^+)\,\theta(k^+) \,,\nn
\end{eqnarray}
where $\omega = \bn\mcdot p$ is the large partonic momentum.

For convenience we define
\begin{eqnarray} \label{defnF0}
  F(p^+,p^-) & = & \int_{0}^{p^+} \! \! dk^+\:
   {\cal J}^{(0)}(p^-, k^+,\mu )\:
   f^{(0)}(k^+  \!+\!\overline\Lambda\!-\! p^+,\mu)  \\
&&
   + \frac{1}{2m_B}\, f_0^{(2)}(\bar\Lambda-p^+)
   - \frac{\lambda_1+3\lambda_2}{2m_B} f^{(0)\,\prime}(\bar\Lambda-p^+)\,,
\nn \\
  F_{1,2}(p^+) & = & f_{1,2}^{(2)}(\bar\Lambda-p^+)   \,, \nn
\end{eqnarray}
where a prime denotes a derivative,
as well as
\begin{eqnarray}
 F_{3,4}(p^+) \!\!& = &\!\! \int\! dk_1^+ \, dk_2^+
  \left[\frac{\delta(k_1^+)-\delta(k_2^+)}{k_2^+ - k_1^+} \right]
  f_{3,4}^{(4)}(k_j^+ +\bar\Lambda-p^+ )\,, \\
 F_{5,6}(p^+) \!\! & = &\!\! \int\! dk_1^+ \, dk_2^+ \, dk_3^+
  \left[\frac{\delta(k_1^+)}{(k_2^+)(k_3^+)}
      \!+\! \frac{\delta(k_2^+)}{(k_1^+)(k_3^+)}
      \!+\! \frac{\delta(k_3^+)}{(k_1^+)(k_2^+)}
      \!-\! \pi^2 \delta(k_1^+)\delta(k_2^+)\delta(k_3^+)
      \right]\nn\\
 && \quad\times
  f_{5,6}^{(6)}(k_j^+ + \bar\Lambda-p^+) \,. \nn
\end{eqnarray}
If we use the tree-level expression for ${\cal J}^{(0)}$, then
$F(p^+,p^-)=F(p^+)$ is a function of $p^+$ only.
Then, for $B\to X_s\gamma$, the rate $d\Gamma^s/dx_H^\gamma$ in
the endpoint region is \cite{klis04}\footnote{This includes 
${\cal O}_7$ -- ${\cal O}_7$ and ${\cal O}_7$ -- ${\cal O}_2$ 
contributions only. In Ref.~\cite{Lee:2006wn}
subleading corrections from ${\cal O}_7$ -- ${\cal O}_8$ are 
studied and estimated to contribute between $-0.3\%$ and $-3\%$
to the total flavour-averaged decay rate. We do not consider such
corrections in this work.}
\begin{eqnarray} \label{eq:dGsdx}
 \frac{1}{\Gamma_0^s} \frac{d\Gamma^s}{dx_H^\gamma}\bigg|_{x_H^\gamma> x_H^c}
 \!\!\! &=&  m_B (C^{(t)})^2 \big[ 1 - 3 (1\!-\!x_H^\gamma)\big]
    F\big(m_B(1\!-\! x_H^\gamma),m_B\big)
  \\
  &&
  + \big[m_B(1\!-\!x_H^\gamma)\!-\!\overline\Lambda\,\big]
    F\big(m_B(1\!-\! x_H^\gamma)\big) 
         + F_2\big(m_B(1\!-\!x_H^\gamma)\big)
\nn\\[4pt]
 & &  
         - F_3\big(m_B(1\!-\!x_H^\gamma)\big)
        +F_4\big(m_B(1\!-\!x_H^\gamma)\big) 
- 8\pi\alpha_s(\mu_i)\ F_5^s\big(m_B(1\!-\!x_H^\gamma)\big)
    \,, \nn
\end{eqnarray}
where $1-x_H^c \sim \lambda^2$ and
\begin{eqnarray}
C^{(t)} & = &  1 + \Delta_\gamma(m_b,\varrho) 
- \frac{\alpha_s(m_b)C_F}{4\pi} \bigg\{ 
  \frac{\pi^2}{12} + 6 \bigg\} \,, \nn \\
\Delta_\gamma(m_b,\varrho) & = & 
\frac{1}{ C_7^{\rm eff(0)}(m_b)} 
 \bigg\{ \frac{\alpha_s}{4\pi} C_7^{\rm eff(1)}(m_b) 
 + \sum_{k} C_k^{\rm eff (0)}(m_b) r_k(\varrho) 
  \bigg\} \,.
\end{eqnarray}
The triply differential decay rate for $B\to X_u \ell\bar\nu$ at NLO
\cite{klis04} is obtained by substituting the $W_i^u$ listed in 
Appendix~\ref{app:hi} into Eq.~(\ref{dGamma3u}). 
At tree level, this becomes
\begin{eqnarray} 
\frac{1}{\Gamma_0^{u}} 
\frac{d^3\Gamma^{u}}{ dx_H\, d\overline y_H\, du_H}  
& = & 
6 (1-u_H) (x_H + \bar y_H - 1) \bigg\{ 2 m_B 
(2 - x_H - \bar y_H - u_H) F(m_B u_H) 
\nn \\
& - & \frac{1}{\bar y_H\!-\!u_H}
\left(\bar y_H^2 - (2\!-\!x_H) \bar y_H + 2 (1\!-\!x_H)
-u_H(2\!-\!x_H\!-\!u_H) \right) \, F_1(m_B u_H)
\nn \\
& + & \frac{2}{\bar y_H (\bar y_H\!-\!u_H)}
\big(\bar y_H^3 - (2\!-\!x_H)\bar y_H^2 - (4\!-\!u_H)(x_H\!+\!u_H)\bar y_H 
 \nn \\
& & \qquad\qquad\qquad
+ 2(x_H\!+\!2\bar y_H\!+\!u_H\!-\!1)\big)
\, F_2(m_B u_H)
\nn \\
& + & \frac{2}{\bar y_H}(x_H+\bar y_H+u_H-2) \, F_3(m_B u_H)
\nn \\
& - & \frac{2}{\bar y_H(\bar y_H\!-\!u_H)}
\left(\bar y_H^2 \!-\! (2\!-\!x_H) \bar y_H 
\!+\! 2(1\!-\!x_H) \!-\! u_H(2\!-\!x_H\!-\!u_H)\right) 
\, F_4(m_B u_H)
\nn \\
& - & \frac{4}{\bar y_H(\bar y_H\!-\!u_H)}(1-\bar y_H)(x_H + \bar y_H - 1) 
\, 4\pi\alpha_s(\mu_i) F_5^u(m_B u_H)
\nn \\
& + & \frac{4}{\bar y_H(\bar y_H\!-\!u_H)}(1-u_H)(1-x_H-u_H) \, 
4\pi\alpha_s(\mu_i) F_6^u(m_B u_H) \bigg\} \,. 
   \label{eq:d3Gexp}
\end{eqnarray}
Note that we can use the relation
\cite{blm1}
\begin{align}
F_1(m_B u_H) &= 2 (\bar\Lambda - m_B u_H) F(m_B u_H) + {\cal O}(\lambda^4)
   \label{eq:F1}
\end{align}
to eliminate $F_1(m_B u_H)$,
as was done in Eq.~(\ref{eq:dGsdx}).

The triply differential decay rate for $B\to X_s\ell^+\ell^-$
was calculated in Refs.~\cite{Lee:2005pk,klft07}. 
The $W_i^{\ell\ell}$ appearing in Eq.~(\ref{dGamma3u}) are also listed
in Appendix~\ref{app:hi}.

\section{$|V_{ub}|$ at NLO} \label{shape}

\subsection{Relations between $B\to X_u\ell\bar\nu$ and $B\to X_s\gamma$}

Consider first the process $B\to X_u\ell\bar\nu$.
We wish to isolate or eliminate the subleading shape functions
that appear in the rates.
In the following, we shall work at tree level.
Inspection of Eqs.~(\ref{hiuLOf}) and (\ref{eq:NLOhu13}) shows that the 
shape functions appear in the hadronic structure functions $W_1$ to $W_3$ 
in only two combinations, namely
\begin{eqnarray}
m_B {\cal F}_I & = & m_B F + \frac{1}{2} F_1 -  F_2 
   -  \frac{1}{\bar y_H}\left( F_3 - F_4 
   +  8\pi\alpha_s(\mu_i) F_5^u \right), \nn\\
m_B {\cal F}_{II} & = & F_1 
   - \frac{2\left(\bar y_H(2-u_H) - 1\right)}{\bar y_H(1-u_H)} F_2
   + \frac{2}{\bar y_H} \left( F_4 - 4\pi\alpha_s(\mu_i) F_5^u 
                                   - 4\pi\alpha_s(\mu_i) F_6^u \right),
\end{eqnarray}
where we have suppressed the argument $m_B u_H$.
Specifically,
\begin{eqnarray}
W_1 & = & \frac{1}{4} {\cal F}_I \,, \nn \\
W_2 & = & \frac{1-u_H}{\bar y_H-u_H} {\cal F}_I 
     - \frac{(1-u_H)^2}{(\bar y_H-u_H)^2} {\cal F}_{II} \,, \nn \\
W_3 & = & \frac{1}{2 m_B (\bar y_H - u_H)} {\cal F}_I \,. 
\end{eqnarray}
Nevertheless, taking integrals of the form
\begin{eqnarray}
\int_{u_H}^{1} d\bar y_H \int_{1-\bar y_H}^{1-u_H} d x_H
K^u(x_H, \bar y_H, u_H) \frac{d^3\Gamma^u}{d x_H d\bar y_H d u_H} \,,
   \label{eq:Ku}
\end{eqnarray}
with suitable choices of the weight function $K^u(x_H, \bar y_H, u_H)$,
we can isolate the following   
four linearly independent combinations of the $F_i$:
\begin{subequations} \label{combs}
\begin{align} 
& (4-2u_H) m_B F + F_1 \; , \label{comba} \\[2pt]
& (1-u_H) m_B F + F_2 \; , \label{combb} \\[2pt]
& F_3 - F_4 + 8 \pi\alpha_s(\mu_i) F_5^u \; , \label{combc} \\ 
& m_B F - \frac{1}{2} F_3 - \frac{1}{2} F_4 + 4\pi\alpha_s(\mu_i) F_6^u \,.
\label{combd}
\end{align}
\end{subequations}
(Recall that we can apply Eq.~(\ref{eq:F1}) so that the first combination 
involves only the leading-order shape function.)
Here, the treatment of the $u_H$ dependence in the rate requires care. 
Expanding Eq.~(\ref{eq:d3Gexp}) in $u_H \sim \lambda^2$ when obtaining
the weight function will typically result in excessively large coefficients 
in the $u_H F_{1-6}(m_B u_H)$ terms (which are formally of order $\lambda^4$).
For example, choosing 
$K^u(x_H, \bar y_H) = -21 x_H + 21 \bar y_H + 45 x_H \bar y_H
- \frac{75}{2} \bar y_H^2$,
we obtain
\begin{eqnarray}
\frac{1}{\Gamma_0^u}\iint d x_H d\bar y_H K^u(x_H, \bar y_H) 
\frac{d^3\Gamma^u}{d x_H d\bar y_H d u_H} 
& = & 
(1 - 7 u_H) m_B F(m_B u_H) + \frac{1}{4} F_1(m_B u_H) 
+ {\cal O}(\lambda^4) \,, \nn \\
\end{eqnarray}
so this eliminates all but the leading-order shape function up to
${\cal O}(\lambda^4)$ corrections.
However, we then have the additional contributions
\begin{align}
& \frac{5}{4} u_H F_1(m_B u_H) - \frac{49}{2} u_H  F_2(m_B u_H)
- \frac{109}{4} u_H F_3(m_B u_H) + \frac{57}{4} u_H F_4(m_B u_H)
\nn \\
& - \frac{83}{2} u_H \times 4\pi\alpha_s(\mu_i) F_5^u(m_B u_H)
+ 13  u_H \times 4\pi\alpha_s(\mu_i) F_6^u(m_B u_H)
\,. 
\end{align}
For this reason, when calculating $K^u$, we keep the 
full dependence on $u_H$ in the rate, 
rather than dropping terms that are formally 
subleading in a strict SCET expansion in $u_H/\bar y_H \sim \lambda^2$. 
(The analysis of $m_X$-cut effects in $B\to X_s\ell^+\ell^-$ 
\cite{Lee:2005pk,Lee:2005pw} also retained the full $u_H$ dependence, 
since doing so facilitates making contact with the total rate in the 
local OPE~\cite{Mannel:2004as,frank,bjorn}.) Thus, subleading shape 
functions are eliminated to all orders in $u_H$, and the issue is resolved.
One straightforward method for obtaining $K^u(x_H,\bar y_H, u_H)$ is
then to take different moments of the rate with respect to $x_H$ and 
$\bar y_H$, and solve the resulting set of linear equations in the $F_i$. 
In Eq.~(\ref{eq:Ku}), we consider the case where a cut is imposed 
on $p_X^+$, i.e.\ $p_X^+ < m_D^2/m_B$.
Different or additional cuts will change the limits of integration,
calling for different weight functions.
Table~\ref{table:01} lists several examples of $K^u$s that isolate
the combination $m_B F + F_1/(4-2u_H)$, while Tables~\ref{table:02}
and \ref{table:345} give examples that result in 
(\ref{combb}) and (\ref{combc}) respectively.



\begin{table}[htb!]
\fbox{
\begin{minipage}{12cm}
{\normalsize
\begin{align}
(1)\qquad K^u_{\rm I} &= \frac{5}{9}\frac{1}{(2-u_H)}\frac{1}{(1-u_H)^8}
\Big[	10(7-u_H)(1-u_H)(4+3u_H) \bar y_H \nn \\
& \qquad - (454+247u_H-71u_H^2) x_H \bar y_H 
	- 4(1-u_H)(109-4u_H) \bar y_H^2 \nn \\
& \qquad + 105(7-u_H) x_H \bar y_H^2
\Big]
\nn \\
(2)\qquad K^u_{\rm I} &= \frac{5}{32}\frac{1}{(2-u_H)}\frac{1}{(1-u_H)^8}
\Big[	-\!10(7-u_H)(1-u_H)(34-27u_H) \bar y_H \nn \\
& \qquad + 2(1-u_H)(2759-449u_H) x_H \bar y_H 
	- 525(7-u_H) x_H^2 \bar y_H \nn \\
& \qquad + 2(1-u_H)(341-131u_H) \bar y_H^2
\Big]
\nn \\
(3)\qquad K^u_{\rm I} &= \frac{15}{41}\frac{1}{(2-u_H)}\frac{1}{(1-u_H)^8}
\Big[	-\!2(1-u_H)^2(288-29u_H) \bar y_H \nn \\
& \qquad + (1426-1793u_H+157u_H^2) x_H \bar y_H 
	- 10(109-4u_H) x_H^2\bar y_H \nn \\
& \qquad + (341-131u_H) x_H \bar y_H^2
\Big]
\nn 
\end{align}
}
\end{minipage}
}
\caption{Some choices of $K^u(x_H,\bar y_H,u_H)$ for which 
the weighted integral Eq.~(\ref{eq:Ku}) 
equals $m_B F + F_1/(4-2u_H)$.
}
   \label{table:01}
\end{table}

 
%
\medskip
\begin{table}[htb!]
\fbox{
\begin{minipage}{12cm}
{\normalsize
\begin{align}
({\rm A})\qquad K^u_{\rm I\!I} &= \frac{5}{9}\frac{1}{(1-u_H)^7}
\Big[	-\!2(1-u_H)(7-15u_H) \bar y_H 
 - (34+71u_H) x_H \bar y_H \nn \\
& \qquad\qquad - 16(1-u_H) \bar y_H^2 
 + 105 x_H \bar y_H^2
\Big]
\nn \\
({\rm B})\qquad K^u_{\rm I\!I} &= \frac{5}{32}\frac{1}{(1-u_H)^7}
\Big[	-\!2(1-u_H)(266-135u_H) \bar y_H 
 + 898(1-u_H) x_H \bar y_H \nn \\
& \qquad\qquad - 525 x_H^2 \bar y_H 
 + 262(1-u_H) \bar y_H^2
\Big]
\nn \\
({\rm C})\qquad K^u_{\rm I\!I} &= \frac{15}{41}\frac{1}{(1-u_H)^7}
\Big[	-\!58(1-u_H)^2 \bar y_H 
 + (26-157u_H) x_H \bar y_H \nn \\
& \qquad\qquad - 40 x_H^2\bar y_H 
 + 131 x_H \bar y_H^2
\Big]
\nn 
\end{align}
}
\end{minipage}
}
\caption{Some choices of $K^u(x_H,\bar y_H,u_H)$ for which 
the weighted integral Eq.~(\ref{eq:Ku}) equals 
$(1\!-\!u_H)m_BF + F_2$.}
   \label{table:02}
\end{table}

\begin{table}[htb!]
\fbox{
\begin{minipage}{12cm}
{\normalsize
\begin{align}
({\rm a})\qquad K^u_{\rm I\!I\!I} &= -\frac{10}{9}\frac{1}{(1-u_H)^8}
\Big[  -\!2(1-u_H)(58+32u_H+15u_H^2) \bar y_H \nn \\
& \qquad\ + (158+104u_H+53u_H^2) x_H \bar y_H 
	+ (1-u_H)(149+61u_H) \bar y_H^2 \nn \\
& \qquad - 105(2+u_H) x_H \bar y_H^2
\Big]
\nn \\
({\rm b})\qquad K^u_{\rm I\!I\!I} &= -\frac{15}{16}\frac{1}{(1-u_H)^8}
\Big[	2(1-u_H)(92-12u_H-45u_H^2) \bar y_H \nn \\
& \qquad - 2(1-u_H)(246+139u_H) x_H \bar y_H 
	+ 175(2+u_H) x_H^2 \bar y_H \nn \\
& \qquad -2(1-u_H)(18+17u_H) \bar y_H^2
\Big]
\nn \\
({\rm c})\qquad K^u_{\rm I\!I\!I} &= -\frac{15}{41}\frac{1}{(1-u_H)^8}
\Big[	2(1-u_H)^2(166+93u_H) \bar y_H \nn \\
& \qquad - 2(483-320u_H-268u_H^2) x_H \bar y_H 
	+ 5(149+61u_H) x_H^2\bar y_H \nn \\
& \qquad - 6(18+17u_H) x_H \bar y_H^2
\Big]
\nn 
\end{align}
}
\end{minipage}
}
\caption{Some choices of $K^u(x_H,\bar y_H,u_H)$ for which 
the weighted integral Eq.~(\ref{eq:Ku}) equals 
$F_3 - F_4 + 2\tilde F_5^u$.}
   \label{table:345}
\end{table}

Now, the subleading shape functions $F_{5,6}$ depend upon the light-quark 
flavour (see Appendix~\ref{app:sf}). We indicate this difference between 
the $F_{5,6}$'s  appearing in $B\to X_u\ell\bar\nu$  and $B\to X_s\gamma$ 
by using the superscripts `u' and `s'.
In order to cancel the $F_5^s$ contribution to the latter decay,%
\footnote{The authors of Refs.~\cite{Bosch,Beneke} have used 
model-dependent arguments 
to estimate that the effects of $f_{5,6}$, when integrated over a
sufficiently large region, are comparatively small ($\sim 5\%$), 
but that they may cause large corrections in the $d\Gamma/dp_X^+$
spectrum for $p_X^+ \leq 0.5\,\mbox{GeV}$. 
We avoid any need to consider the reliability of these numerics
by simply eliminating $f_{5,6}$, along with the other tree-level 
shape functions.}
we can use approximate $SU(3)$ flavour symmetry,
namely 
the fact that
\begin{align}
\frac{F_5^u - F_5^s}{F_5^s} &\sim \frac{m_s}{\Lambda_{\rm QCD}}
\end{align}
is suppressed.
This enables us to relate the semileptonic process to the radiative
process and thereby derive an expression for $\Gamma_0^u$, or
equivalently $|V_{ub}|$, to subleading order.
We can write 
\begin{align}  \label{eq:bumom}
\lefteqn{
\frac{1}{\Gamma_0^u} \iint
\Big[K^u_{\rm I\!I} - K^u_{\rm I\!I\!I} + \rho K^u_{\rm I}\Big] 
\frac{d^3\Gamma^u}{d x_H d\bar y_H d u_H} d x_H d\bar y_H 
} \\
&= m_B F(m_B u_H) - \frac{1}{2} \big[ F_1(m_B u_H)
    - 2 F_2(m_B u_H) \big]
 \nn \\
& \qquad 
   - \big[ F_3(m_B u_H)
   - F_4(m_B u_H) + 2 \tilde F_5^u(m_B u_H) \big]  \,, \nn
\end{align}
where
\begin{align}
\rho(u_H) &= \frac{(2-u_H)\big(u_H-\frac{\bar\Lambda}{2m_B}\big)}
{\big(1-u_H +\frac{\bar\Lambda}{2m_B}\big)}\,
\end{align}
and $\tilde F_5^u = 4\pi\alpha_s(\mu_i) F_5^u$.
$K^u_{\rm I}$, $K^u_{\rm I\!I}$ and $K^u_{\rm I\!I\!I}$ are any 
weight functions that give the combinations $m_B F + F_1/(4\!-\!2u_H)$,
$(1\!-\!u_H)m_BF + F_2$ and $F_3-F_4 + 2\tilde F_5^u$
respectively (examples of which are presented in Tables~\ref{table:01},
\ref{table:02} and \ref{table:345}).
The shape functions in Eq.~(\ref{eq:bumom}) appear in the same linear
combination as in the rate $d\Gamma^s/du_H$.
Hence, at NLO we obtain 
\begin{align}
\frac{1}{\Gamma_0^u} 
\iint
\Big[K^u_{\rm I\!I} - K^u_{\rm I\!I\!I} +\rho K^u_{\rm I} \Big]
\frac{d^3\Gamma^u}{d x_H d\bar y_H d u_H} d x_H d\bar y_H
&=-\frac{1}{(1\!-\!u_H)^3}\,\frac{1}{\Gamma_0^s} \frac{d \Gamma^s}{d u_H} \,.
   \label{eq:res0}
\end{align}

More generally, we can construct $K^u$ such that
\begin{align}
\hat M^u &\equiv \frac{1}{\Gamma_0^u} M^u
\equiv  \frac{1}{\Gamma_0^u} \iint
K^u(x_H, \bar y_H, u_H) \frac{d^3\Gamma^u}{d x_H d\bar y_H d u_H}
d x_H d\bar y_H  \\
  &= m_B F(m_B u_H) + \kappa_1^u(u_H) F_1(m_B u_H) \nn \\
  & \qquad + \kappa_2^u(u_H) \big[ F_2(m_B u_H) - F_3(m_B u_H)
   + F_4(m_B u_H) - 2 \tilde F_5^u(m_B u_H) \big]  \,. \nn 
\end{align}
For example, we can use
\begin{equation}
K^u = K^u_{\rm I\!V} - \kappa_2^u K^u_{\rm I\!I\!I}, 
   \label{eq:egK0}
\end{equation}
where 
$K^u_{\rm I\!V}$ is a weight function that gives the linear combination
$m_B F + \kappa_1^u F_1 + \kappa_2^u F_2$,
examples of which are given in Table~\ref{table:012} in 
Appendix~\ref{app:Ku} 
(with the corresponding values of $\kappa_1^u$ and $\kappa_2^u$ shown
there). 
We can also use
\begin{align}
K^u = \beta K^u_{\rm I} + \frac{1-\beta}{1-u_H} (K^u_{\rm I\!I} 
- K^u_{\rm I\!I\!I})\,,
   \label{eq:egK}
\end{align}
with $\beta$ an arbitrary real number
(in which case $\kappa_1^u = \beta/(4-2u_H)$ and 
$\kappa_2^u = (1-\beta)/(1-u_H)$).
For any such $K^u$, we have
\begin{align} \label{eq:Mc}
\hat M^u + \kappa_2^u (1\!-\!u_H)^{-3} \hat M^s
&= \left\{ \left( 1 - \kappa_2^u\right) + 
\left(\frac{\bar\Lambda}{m_B}-u_H\right)
\left(2 \kappa_1^u + \kappa_2^u\right)\right\} m_B F(m_B u_H)
 \\[3pt]
& \qquad\qquad 
+ {\cal O}(\alpha_s,\lambda^4) \,,
\nn
\end{align}
where 
$\hat M^s =(1/\Gamma_0^s) M^s = (1/\Gamma_0^s) (d\Gamma^s/du_H)$,
i.e.\ combining $\hat M^u$ and $\hat M^s$ in this
way gives an expression dependent only on the leading-order shape
function.  Taking the ratio of two such expressions (two choices of
$K^u$) at $u_H \neq 0$ then provides us with a relation
independent of both leading and subleading shape functions.
We shall use the superscripts $(i)$ and $(ii)$ when we need to
distinguish between quantities in the two expressions.
We then obtain
\begin{align}
\frac{\Gamma_0^u}{\Gamma_0^s} &= 
- \frac{\left[ b_0^{(ii)} M^{u(i)} - b_0^{(i)}M^{u(ii)} \right]}
{\left[ b_0^{(ii)} \kappa_2^{u(i)} - b_0^{(i)} \kappa_2^{u(ii)}\right]
(1\!-\!u_H)^{-3} M^s}\,,
   \label{eq:res}
\end{align}
where
\begin{align}
b_0 &= \left( 1 - \kappa_2^u\right) +
\left(\frac{\bar\Lambda}{m_B}-u_H\right)
\left(2 \kappa_1^u + \kappa_2^u\right) \,.
\end{align}
Since the right-hand side of Eq.~(\ref{eq:res}) is measurable,
it enables an experimental determination of the CKM ratio on 
the left-hand side.
Additionally, the factor $|V_{tb} V_{ts}^*|$ in this ratio can be
eliminated by normalizing the photon spectrum by the total 
$B\to X_s\gamma$ rate, which is given in a local OPE.

There will be loop and power ($\lambda^4$-suppressed) corrections to
the rates and hence also to Eq.~(\ref{eq:res}).
While these are not fully known, one can show that the corrections
to Eq.~(\ref{eq:res}) are proportional to
\begin{align}
-\frac{b_0^{(i)}}{b_0^{(ii)} - b_0^{(i)}}
+\frac{b_0^{(i)} \kappa_2^{u(ii)}}
{b_0^{(ii)} \kappa_2^{u(i)}
- b_0^{(i)} \kappa_2^{u(ii)}}
 + \cdots
\,
  \label{correct}
\end{align}
(multiplied by $\alpha_s$ or $\lambda^4$).
This needs to be taken into account when selecting
$\{ K^{u(i)}, K^{u(ii)} \}$:
one should avoid pairs of weight functions that result in
Eq.~(\ref{correct}) being excessively large, lest parametrically
suppressed terms acquire excessively large coefficients.
For example, one appropriate choice is to use Eq.~(\ref{eq:egK}) for both 
$K^u$s, with $\beta^{(i)}=1$ and $\beta^{(ii)}=0$, after which
the magnitude of Eq.~(\ref{correct}) is less than $1/6$ for 
$0<u_H < m_D^2/m_B^2$.


\subsection{Relations involving $B\to X_s\ell^+\ell^-$}

We can also try to isolate shape functions in the process 
$B\to X_s\ell^+\ell^-$ by taking integrals of the form
\begin{eqnarray}
\int_{\bar y_{\rm min}}^{\bar y_{\rm max}} d\bar y_H 
\int_{1-\bar y_H}^{1-u_H} d x_H
K^{\ell\ell}(x_H, \bar y_H, u_H) 
\frac{d^3\Gamma^{\ell\ell}}{d x_H d\bar y_H d u_H} \,,
   \label{eq:Kll}
\end{eqnarray}
where 
\begin{align}
\bar y_{\rm min(max)} &= 1 - \frac{y_H^{\rm max(min)}}{(1-u_H)} \,.
\end{align}
Here, $y_H = q^2/m_B^2$ and the low-$q^2$ region corresponds to
$1\, \mbox{GeV}^2 \leq q^2 \leq 6 \,\mbox{GeV}^2$.
However, determining $K^{\ell\ell}(x_H, \bar y_H, u_H)$
in the straightforward manner described above proves to be problematic 
in practice.
Therefore, we resort to another method, which is based on
the following observation. Under the transformation
$ x_H \to x_H' = 2\!-\!u_H\!-\!\bar y_H\!-\!x_H$, we find that 
$\int_{1-\bar y_H}^{1-u_H} dx_H = \int_{1-\bar y_H}^{1-u_H} dx_H'$
and
\begin{eqnarray*}
(1-x_H -u_H) & \leftrightarrow & (x_H + \bar y_H - 1) \,, \\
(2 x_H + u_H + \bar y_H - 2) & \leftrightarrow & - (2 x_H + u_H + \bar y_H - 2)
\,.
\end{eqnarray*}
This symmetry or antisymmetry can be exploited to obtain $K^{\ell\ell}$. 
For example, if $K^{\ell\ell}$ changes sign under the transformation,
then we can see from the triply differential rate, Eq.~(\ref{dGamma3u}),
that integration over $x_H$ eliminates the $W_1$ and $W_2$ terms, whereas 
the $W_3$ term remains. Now, Eq.~(\ref{eq:NLOhll}) shows that 
$F_3$, $F_4$ and $F_5^s$ occur in $W_3$ in the same linear combination as 
in the $B\to X_s\gamma$ rate.

This still leaves the integration over $\bar y_H$, and if we choose
$K^{\ell\ell}(x_H,\bar y_H,u_H) = (2 x_H + u_H + \bar y_H - 2) \,
\tilde K^{\ell\ell}(\bar y_H,u_H)$, where 
$\tilde K^{\ell\ell}(\bar y_H,u_H)$ satisfies
\begin{equation}
\int_{\bar y_{\rm min}}^{\bar y_{\rm max}} 
d\bar y_H (\bar y_H\!-\!u_H)^3 \frac{1}{\bar y_H}
\big( 2 \mbox{Re}[ C_{10a}\, C_{7a}^{\,*}] + 
\mbox{Re}[C_{10a}\, C_{9a}^{\,*}](1-\bar y_H^2) 
\big) \tilde K^{\ell\ell}(\bar y_H,u_H) = 0 \,,
\end{equation}
then all of the subleading shape functions in Eq.~(\ref{eq:Kll})
appear in the same combination as in the $B\to X_s\gamma$ rate, 
which can thus be used to eliminate these functions.
Table~\ref{table:ll23} in Appendix~\ref{app:Ku} shows several examples of 
$K^{\ell\ell}$ of this form. %
%
We observe that  
$z= \cos\theta = (2x_H + u_H + \bar y_H -2)/(\bar y_H-u_H)$,
where $\theta$ is the angle between the $B$ and
$\ell^+$ in the center-of-mass frame of the $\ell^+\ell^-$.
This means that a choice of $K^{\ell\ell} \propto
(2x_H + u_H + \bar y_H -2)$ is equivalent to taking moments of the 
forward-backward asymmetry, 
\begin{align}
\frac{d^2 A_{FB}}{d\bar y_H du_H} 
&
= \int_{-1}^{1}dz \frac{\mbox{sign}(z)}{\Gamma_0}
\frac{d^3\Gamma}{d\bar y_H du_Hdz} 
\, = \, \frac{3}{2\Gamma_0} \int_{-1}^{1}dz \, z \,
\frac{d^3\Gamma}{d\bar y_H du_Hdz} \,.
\end{align}
Note also that $C_{9a}$ is a function of $q^2$, and hence of $\bar y_H$
(see Appendix~\ref{app:hi}),
but in the low-$q^2$ region $|C_{9a}|$ varies by less than $\pm 1\%$
and we take it to be constant. There is no problem taking into
account the exact dependence, but integrals over regions of $\bar y_H$ 
must then be performed numerically.

Let 
$\Gamma_0^s \hat M^s = d\Gamma^s/du_H$, and let 
$M^u = \Gamma_0^u \hat M^u$ and $\Gamma_0^{\ell\ell} \hat M^{\ell\ell}$
denote the integrals (\ref{eq:Ku}) and (\ref{eq:Kll}) respectively, with 
weight functions from Tables~\ref{table:01} and \ref{table:ll23}. 
Then we obtain
\begin{align}
\Gamma_0^u &= 
\frac{1+\kappa_3^{\ell\ell}}
{1+2\big(\frac{\bar \Lambda}{m_B}-u\big)
\kappa_1^u} 
\;
\frac{M^u}{\hat M^{\ell\ell}
- \kappa_3^{\ell\ell}(1-u_H)^{-3} \hat M^s} \,,
   \label{result0}
\end{align}
where $\kappa_1^u$ ($\kappa_3^{\ell\ell}$) is the coefficient of
$F_1$ ($F_3$) in $\hat M^u$ ($\hat M^{\ell\ell}$).

More generally, by the same methods, we can find
$K^u$ and $K^{\ell\ell}$ such that
\begin{align} \label{eq:Musll}
\hat M^u &\equiv \frac{1}{\Gamma_0^u} M^u 
\equiv  \frac{1}{\Gamma_0^u} \iint
K^u(x_H, \bar y_H, u_H) \frac{d^3\Gamma^u}{d x_H d\bar y_H d u_H} 
d x_H d\bar y_H  \\
  &= m_B F(m_B u_H) + \kappa_1^u(u_H) F_1(m_B u_H) 
     + \kappa_2^u(u_H) F_2(m_B u_H)  \,, \nn \\
\hat M^s &\equiv  \frac{1}{\Gamma_0^s} M^s
\equiv \frac{1}{\Gamma_0^s} \frac{d \Gamma^s}{d u_H} \nn \\
&= - (1-u_H)^3 \left\{ m_B F(m_B u_H) - \frac{1}{2} \big[ F_1(m_B u_H)
    - 2 F_2(m_B u_H) \big] 
\right. \nn \\
& \qquad \left.
   - \big[ F_3(m_B u_H)
   - F_4(m_B u_H) + 2 \tilde F_5^s(m_B u_H) \big] \right\} \,, \nn \\
\hat M^{\ell\ell} &\equiv \frac{1}{\Gamma_0^{\ell\ell}} M^{\ell\ell} 
\equiv \frac{1}{\Gamma_0^{\ell\ell}} \iint K^{\ell\ell}(x_H, \bar y_H, u_H) 
\frac{d^3\Gamma^{\ell\ell}}{d x_H d\bar y_H d u_H} 
d x_H d\bar y_H \nn \\
  &= m_B F(m_B u_H) + \frac{1}{2}\kappa_2^{\ell\ell}(u_H) \big[ F_1(m_B u_H) 
   - 2 F_2(m_B u_H) \big]  
\nn \\
& \qquad
   + \kappa_3^{\ell\ell}(u_H) \big[ F_3(m_B u_H)  
   - F_4(m_B u_H) + 2 \tilde F_5^s(m_B u_H) \big] \,, \nn 
\end{align}
where $\tilde F_5^s = 4\pi\alpha_s(\mu_i) F_5^s$.
Tables~\ref{table:012} and \ref{table:ll} show (further) examples of such 
weight functions, along with the corresponding values of the coefficients 
$\kappa_{1,2}^u$ and $\kappa_{2,3}^{\ell\ell}$.
%
%
%
Then
\begin{align} \label{eq:Mcomb}
\lefteqn{
\big[ \kappa_2^{\ell\ell} - \kappa_3^{\ell\ell} \big]  \hat M^u
+ \kappa_2^u \hat M^{\ell\ell} 
- \kappa_2^u \kappa_3^{\ell\ell} (1-u_H)^{-3} \hat M^s
}
 \\
&= \left\{ \left( \kappa_2^{\ell\ell} - \kappa_3^{\ell\ell} + \kappa_2^u
+ \kappa_2^u \kappa_3^{\ell\ell} \right) 
+ \left(\frac{\bar\Lambda}{m_B}-u_H\right)
\left(\kappa_2^{\ell\ell} - \kappa_3^{\ell\ell}\right) 
\left(2 \kappa_1^u + \kappa_2^u\right)\right\} m_B F(m_B u_H)
\nn \\
& \qquad + {\cal O}(\alpha_s,\lambda^4) \,,
\nn
\end{align}
so in this case we have a combination of $\hat M^u$, $\hat M^s$ and 
$\hat M^{\ell\ell}$ that is dependent only on the leading-order shape 
function.  Taking the ratio of two such expressions (two choices of 
$\{ K^u, K^{\ell\ell} \}$, denoted by superscripts $(i)$ and $(ii)$
as previously) at $u_H \neq 0$ then provides us with another relation 
independent of both leading and subleading shape functions.

Specifically, let
\begin{align} \label{defn1}
\hat{\cal M}^u & = \frac{1}{\Gamma_0^u} {\cal M}^u = \left\{
\begin{tabular}{ll}
$[ \kappa_2^{\ell\ell}-\kappa_3^{\ell\ell} ] \hat M^u$, 
& if $\kappa_2^u \neq 0$ \\
$\hat M^u$, & if $\kappa_2^u = 0$
\end{tabular}
\right.\,\,\,,\,\,\,
\\[2pt]
\hat{\cal M}^{\ell\ell} 
& = \frac{1}{\Gamma_0^{\ell\ell}} {\cal M}^{\ell\ell} 
= \left\{
\begin{tabular}{ll}
$\kappa_2^u \hat M^{\ell\ell}$, 
& if $\kappa_2^{\ell\ell} \neq \kappa_3^{\ell\ell}$ \\
$\hat M^{\ell\ell}$, & if $\kappa_2^{\ell\ell} = \kappa_3^{\ell\ell}$
\end{tabular}
\right.\,\,\,, \nn \\[2pt]
\hat{\cal M}^s & = \frac{1}{\Gamma_0^s} {\cal M}^s = \left\{
\begin{tabular}{ll}
$-\kappa_2^u \kappa_3^{\ell\ell} (1-u_H)^{-3} \hat M^s$, 
& if $\kappa_2^{\ell\ell} \neq \kappa_3^{\ell\ell}$ \\
$-\kappa_3^{\ell\ell} (1-u_H)^{-3} \hat M^s$, 
& if $\kappa_2^{\ell\ell} = \kappa_3^{\ell\ell}$
\end{tabular}
\right.\,\,\,, \nn 
\end{align}
and
\begin{align}
c_0 & = \left\{
\begin{tabular}{ll}
$ \left( \kappa_2^{\ell\ell} - \kappa_3^{\ell\ell} + \kappa_2^u
+ \kappa_2^u \kappa_3^{\ell\ell} \right) 
+ \left(\frac{\bar\Lambda}{m_B}-u_H\right)
\left(\kappa_2^{\ell\ell} - \kappa_3^{\ell\ell}\right) 
\left(2 \kappa_1^u + \kappa_2^u\right) $,
& if $\kappa_2^u \neq 0$  and $\kappa_2^{\ell\ell}\neq\kappa_3^{\ell\ell}$\\
$1+2\left(\frac{\bar\Lambda}{m_B}-u_H\right)\kappa_1^u$, 
& if $\kappa_2^u=0$ \\
$(1+\kappa_3^{\ell\ell})$, & if $\kappa_2^{\ell\ell}=\kappa_3^{\ell\ell}$ 
   \label{defn2}
\end{tabular}
\right.\,\,\,.
\end{align}
We find that
\begin{align}
\frac{\Gamma_0^u}{\Gamma_0^s} &= - \left[
\frac{c_0^{(ii)} {\cal M}^{u(i)} - c_0^{(i)}{\cal M}^{u(ii)}}
{c_0^{(ii)}({\cal M}^{s(i)} + r {\cal M}^{\ell\ell(i)}) 
- c_0^{(i)}({\cal M}^{s(ii)} + r {\cal M}^{\ell\ell(ii)})}\right] \,,
  \label{result}
\end{align}
where $r = \Gamma_0^s/\Gamma_0^{\ell\ell}$, or
\begin{align}
\Gamma_0^u &= - \left[
\frac{c_0^{(ii)} {\cal M}^{u(i)} - c_0^{(i)}{\cal M}^{u(ii)}}
{c_0^{(ii)}(\hat{\cal M}^{s(i)} + \hat{\cal M}^{\ell\ell(i)}) 
- c_0^{(i)}(\hat{\cal M}^{s(ii)} + \hat{\cal M}^{\ell\ell(ii)})}\right] \,.
  \label{result2}
\end{align}
In the special case where $\kappa_2^{u(i)} = 0$ and 
$\kappa_2^{\ell\ell(ii)} = \kappa_3^{\ell\ell(ii)}$,
Eq.~(\ref{result2}) reduces to Eq.~(\ref{result0}).

The loop and power ($\lambda^4$-suppressed) corrections
to Eq.~(\ref{result}) can be shown to be proportional to
\begin{align} \label{correctns}
&
-\frac{\tilde c_0^{(i)} [\kappa_2^{\ell\ell}-\kappa_3^{\ell\ell}]^{(ii)}}
{\tilde c_0^{(ii)} [\kappa_2^{\ell\ell}-\kappa_3^{\ell\ell}]^{(i)} 
- \tilde c_0^{(i)} [\kappa_2^{\ell\ell}-\kappa_3^{\ell\ell}]^{(ii)}} 
+\frac{\tilde c_0^{(i)} [\kappa_2^u(1+\kappa_3^{\ell\ell})]^{(ii)}}
{\tilde c_0^{(ii)} [\kappa_2^u(1+\kappa_3^{\ell\ell})]^{(i)} 
- \tilde c_0^{(i)} [\kappa_2^u(1+\kappa_3^{\ell\ell})]^{(ii)}} 
\\
& \qquad\qquad + \cdots 
\,, \nn
\end{align}
where
$\tilde c_0 = 
 ( \kappa_2^{\ell\ell} - \kappa_3^{\ell\ell} + \kappa_2^u
+ \kappa_2^u \kappa_3^{\ell\ell} ) 
+ (\bar\Lambda/m_B-u_H)
(\kappa_2^{\ell\ell} - \kappa_3^{\ell\ell}) 
(2 \kappa_1^u + \kappa_2^u)$.
When selecting 
$\{ K^{u(i)}, K^{\ell\ell(i)} \} $,
$\{ K^{u(ii)}, K^{\ell\ell(ii)} \}$, 
one should avoid those sets of weight functions that result in 
Eq.~(\ref{correctns}) being excessively large.
The following combinations of weight functions are
suitable choices:

\medskip
\begin{tabular}{l}
 $K^{u(i)}=(1),(2)\,\mbox{or}\,(3)$
[Table~\ref{table:01}] and $K^{\ell\ell(ii)}=(7),(8)\,\mbox{or}\,(9)$
[Table~\ref{table:ll23}]; \\[4pt]
$K^{u(i)}=(4),(5)\,\mbox{or}\,(6)$
[Table~\ref{table:012}],
$K^{\ell\ell(i)}=(10),(11)\,\mbox{or}\,(12)$ [Table~\ref{table:ll}], \\
\hspace{1cm} and $K^{\ell\ell(ii)}=(7),(8)\,\mbox{or}\,(9)$
[Table~\ref{table:ll23}].
\end{tabular}

\subsection{Perturbative Corrections}

Let us now consider the feasibility of incorporating perturbative
corrections in our relations.
In Ref.~\cite{klis04}, the complete set of subleading corrections 
(to all orders in $\alpha_s$)
for the triply differential spectrum of $B\to X_u\ell\bar\nu$ was derived.
It was shown that prohibitively many new shape functions appear at order
$\alpha_s \Lambda_{\rm QCD}/m_b$, and hence it is not phenomenologically
viable to work to that order.\footnote{Unless these shape functions 
appear in the rates in only a much smaller number of linear
combinations.} 
However, one may choose to work to order 
$(\alpha_s \lambda^0,\, \alpha_s^0 \lambda^2)$, by including perturbative
corrections to just the leading-power terms.
Recall that there are two perturbative scales, $\mu_b\sim m_b$ (hard) and  
$\mu_i \sim \sqrt{m_b\Lambda_{\rm QCD}}$ (jet).
It is straightforward to take into account the relevant hard corrections.
Including the effect of corrections to the jet function 
${\cal J}^{(0)}$, which is convoluted with the shape function $f^{(0)}$, 
is more involved: one has to ``invert'' a distribution 
(see Eq.~(\ref{J01loop})). An implementation akin to
Refs.~\cite{Leibovich:1999xf,Leibovich:2000ey,Lange:2005qn,Lange}
is left for future work.
Nevertheless, before this is done, we can still use the less direct 
approach mentioned in the introduction, using two instances of 
Eq.~(\ref{eq:Mc}) or (\ref{eq:Mcomb}), with appropriately modified
right-hand sides.
For example, one can extract the leading-order shape function from the 
analogue of Eq.~(\ref{eq:Mcomb}), with $K^{\ell\ell}$ from 
Table~\ref{table:ll23}, and substitute this function into a second choice, 
with $K^u$ from Table~\ref{table:01}.
Finally, we note that the extent to which Eq.~(\ref{eq:res}) or 
(\ref{result2}) 
varies with respect to $u_H$ or different combinations
of the $K^u$s and $K^{\ell\ell}$s will provide a measure of the
effect of $\alpha_s$ and $\lambda^4$ corrections.

\section{Conclusion} \label{conclusion}

In this paper, we have established a method for obtaining 
$|V_{ub}|/|V_{tb} V_{ts}^*|$
that includes ${\cal O}(\Lambda_{\rm QCD}/m_b)$ corrections in
a model-independent way. 
Our approach relies upon a class of relations between the inclusive decays
$B\to X_u\ell\bar\nu$ and $B\to X_s\gamma$ that are valid including
the first-order power corrections 
(see Eqs.~(\ref{eq:res0}) and (\ref{eq:res})). Alternatively, one
can use a separate class of relations involving  $B\to X_s\ell^+\ell^-$ 
(see Eqs.~(\ref{result0}) and (\ref{result2})).  
Experimentally required cuts make shape-function effects important
in these processes.  Their differential decay spectra in the 
shape function region have previously been derived to subleading order 
with the help of the Soft-Collinear Effective Theory. 
These rates involve a number of non-perturbative but universal 
shape functions in different linear combinations. We are able to eliminate 
these sources of hadronic uncertainty by taking suitable weighted integrals 
of the triply differential rates. 
Hence, our results incorporate NLO power corrections while avoiding 
model dependence.
There are many possible weight functions (see e.g. 
Eqs.~(\ref{eq:egK0}) and (\ref{eq:egK})); different choices provide
a consistency check on the determination of $|V_{ub}|$.

\medskip
\acknowledgments 

I wish to thank Iain Stewart for helpful discussions and for comments on 
the manuscript.
I am also grateful for conversations with Antonio Limosani,
Frank Tackmann and Mark Wise.


\appendix
\section{Shape functions} \label{app:sf}
The leading-order shape function is
\begin{align} \label{f0}
  f^{(0)}(\ell^+) &=
  \frac{1}{2}\: \langle \bar B_v | \bar h_v
  \delta(\ell^+\!-\! in\mcdot D)  h_v | \bar B_v \rangle \,,
\end{align}
where $h_v$ is the heavy quark field.
The subleading shape functions are 
\begin{eqnarray} \label{defnfs1}
  \langle \bar B_v | O_0(\ell^+) | \bar B_v\rangle
    &=& f_0^{(2)}(\ell^+) \,,\\[3pt]
  \langle \bar B_v | O_{1}^{\beta}(\ell^+) | \bar B_v\rangle
    &=& \Big(v^\beta \!-\! \frac{n^\beta}{n\mcdot v}\Big) \:
     f_1^{(2)}(\ell^+) \,,\nn\\[3pt]
  \langle \bar B_v | P_{2}^{\beta\lambda}(\ell^+) | \bar B_v\rangle
    &=& \: \epsilon_\perp^{\beta\lambda} \:
     f_2^{(2)}(\ell^+) \,,\nn\\[3pt]
  \langle \bar B_v | O_{3}^{\alpha\beta}(\ell_{1,2}^+) | \bar B_v\rangle
    &=& \: g_\perp^{\alpha\beta} \:
     f_3^{(4)}(\ell_1^+,\ell_2^+) \,,\nn\\[3pt]
  \langle \bar B_v | P_{4\lambda}^{\alpha\beta}(\ell_{1,2}^+) | \bar B_v\rangle
    &=& \: - \epsilon_\perp^{\alpha\beta}
     \Big(v_\lambda \!-\! \frac{n_\lambda}{n\mcdot v}\Big) \:
     f_4^{(4)}(\ell_1^+,\ell_2^+) \,,\nn \\
 n_\alpha n_\beta
 \langle \bar B_v | O_{5}^{\alpha\beta}(\ell_{1,2,3}^+) | \bar B_v\rangle
    &=&  f_5^{(6)}(\ell_1^+,\ell_2^+,\ell_3^+) \:
      \,, \nn \\[3pt]
  (g^\perp_{\alpha\beta}- i \epsilon^\perp_{\alpha\beta})
 \langle \bar B_v | O_{5}^{\alpha\beta}(\ell_{1,2,3}^+) | \bar B_v\rangle
    &=&  f_6^{(6)}(\ell_1^+,\ell_2^+,\ell_3^+) \:
      \,,\nn
\end{eqnarray}
where 
$g_\perp^{\mu\nu}=g^{\mu\nu}-(1/2)(n^{\mu}\bn^{\nu}
+ n^{\nu}\bn^{\mu})$ and
$\epsilon_\perp^{\mu\nu}=(1/2)\epsilon^{\mu\nu\alpha\beta}\bn_\alpha n_\beta$.
The ultrasoft operators are 
\begin{eqnarray}
 O_0^{(2)}(\ell^+) &=&
  \int\!\! \frac{dx^-}{\!8\pi}\: e^{-\frac{i}{2} x^- \ell^+ } \!\!
  \int\!\! d^4y \:
   T\: \big[\bar h_v(\tilde x) Y(\tilde x,0) h_v(0)\ i O_h(y) \big]
  \,,  \\
 O_{1}^{\beta}(\ell^+)\:
 &= & \frac12
 \bar h_v
 \big\{ i D^{\beta}_{\rm us}, \delta(\ell^+ \!\! -\! in\mcdot D_{\rm us})
  \big\} h_v
  \,, \nn \\
  P^\beta_{2\lambda}(\ell^+)
 &=& \frac{i}2 \:
    \bar h_v  \big[  i D^{\beta}_{\rm us},
   \delta(\ell^+ \!\! -\! in\mcdot D_{\rm us})\big]
   \gamma^T_\lambda \gamma_5  h_v
  \,, \nn \\
  O_3^{\alpha\beta}(\ell^+_1,\ell^+_2)
  &=& \frac{1}{2}\:
\bar h_v \delta(\ell_2^+ \!-\! in \mcdot D_{us} )
\left\{iD^{\perp\alpha}_{us}, iD^{\perp\beta}_{us}\right\}\delta
(\ell_1^+ \!-\! i n \mcdot D_{us}) h_v\,,
   \nn \\
P_{4\lambda}^{\alpha\beta}(\ell^+_1,\ell^+_2)
 &=& -\frac{1}2 \:
 \bar h_v \delta(\ell^+_2\!-\! in \mcdot D_{us}  )g G_{us \perp}^{\alpha\beta}
  \delta(\ell^+_1\!-\! i n \mcdot D_{us}  )\gamma_\lambda^T \gamma_5  h_v
  \,, \nn \\
O_5^{\alpha\beta}(\ell^+_{1,2,3})
 &=& \frac{1}{2} \:
 \big\{\bar h_v \delta( \ell^+_3\!-\! in \mcdot D_{us} )\gamma^\beta
   P_L T^A q^\bn\big\}\:
 \delta(\ell_2^+-in\mcdot \partial)\: \big\{\bar q^\bn \gamma^\alpha P_L
  \delta(\ell^+_1\!-\! i n \mcdot D_{us}  ) T^A h_v \big\}\,, \nn 
\end{eqnarray}
where $\tilde x^\mu = \bn\mcdot x \, n^\mu/2$.
Here, $O_h$ is the NLO term in the HQET Lagrangian, $Y$ is an ultrasoft
Wilson line, 
$ig G_{us\perp}^{\mu\nu}=[iD_{us}^{\perp\mu}, iD_{us}^{\perp\nu}]$
and $q_{us}^\bn=(\bnslash\nslash)/4\, q_{us}$.
The operator $O_5^{\alpha\beta}$, which appears in the definitions of 
$f_{5,6}$, depends upon the light-quark flavour, $u$ or $s$.

\section{Hard Coefficients} \label{app:hi}
In this appendix, we present expressions for the hard coefficients
in $B\to X_u\ell\bar\nu$ and $B\to X_s\ell^+\ell^-$
\cite{klis04,Lee:2005pk,klft07}.
At lowest order, we have
\begin{eqnarray} \label{Wfact00}
W_i^{(0)} &=&
  h_i(p_X^-,m_b,\mu) \:
  \int_{0}^{p_X^+} \! \! dk^+\:
   {\cal J}^{(0)}(p_X^-\, k^+,\mu )\:
   f^{(0)}(k^+  \!+\!\overline\Lambda\!-\! p_X^+,\mu) \,.
\end{eqnarray}
For $B\to X_u\ell\bar\nu$, we have 
\begin{eqnarray} \label{hiuLOf}
  h_1^{u} &=& \frac{1}{4} \big[ C_1^{(v)} \big]^2  \,,\qquad\\
  h_2^{u} &=&
   \frac{(1-u_H) \, \big[ (C_1^{(v)})^2 + C_1^{(v)} C_2^{(v)}+
     C_2^{(v)} C_3^{(v)} \big] }{(\overline y_H\! -\! u_H)}
   + \frac{(C_2^{(v)})^2}{4}
   + \frac{(1-u_H)^2\, \big[ (C_3^{(v)})^2 +  2 C_1^{(v)} C_3^{(v)} \big]}
    {(\overline y_H\! -\! u_H )^2}    \,,\nn\\
  h_3^{u} &=& \frac{(C_1^{(v)})^2}{2m_B(\overline y_H \! -\! u_H )} \,,\qquad
\nn
\end{eqnarray}
where 
\begin{eqnarray} \label{matchC}
 C_{1}^{(v)}(\hat\omega,1)
 &=& 1 - \frac{\alpha_s(m_b)C_F}{4\pi} \bigg\{ 2\!\ln^2(\hat\omega)
  + 2 {\rm Li}_2(1\!-\!\hat\omega)
  + \ln(\hat\omega) \Big( \frac{3\hat\omega-2}{1-\hat\omega}\Big)
  + \frac{\pi^2}{12} + 6\bigg\} , \nn \\*
C_{2}^{(v)} (\hat\omega,1)
 &=& \frac{\alpha_s(m_b)C_F}{4\pi}\:  \bigg\{ \frac{2}{(1-\hat\omega)}
  + \frac{2\hat\omega\ln(\hat\omega)} {(1-\hat\omega)^2} \bigg\} \,, \nn \\
C_{3}^{(v)}(\hat\omega,1)
 &=& \frac{\alpha_s(m_b)C_F}{4\pi} \bigg\{
  \frac{(1-2\hat\omega)\hat\omega \ln(\hat\omega) }{(1-\hat\omega)^2}
  - \frac{\hat\omega}{1-\hat\omega} \bigg\}
 \,. 
\end{eqnarray}
Here, $\hat\omega=\omega/m_b$. 
%

For $B\to X_s\ell^+\ell^-$, we have
\begin{eqnarray} \label{hillLO}
  h_1^{\ell\ell} &=&
   \frac{1}{2} \Big( \big|{\cal C}_{9}\big|^2
     + \big|{\cal C}_{10a}\big|^2 \Big) +
  \frac{2\,
    \mbox{Re}\big[ {\cal C}_{7}\, {\cal C}_{9}^{\,*} \big]
  }{(1\!-\!\bar y_H)}\
  +
  \frac{2\,\big| {\cal C}_{7} \big|^2 }{(1\!-\!\bar y_H)^2}\,, \\[5pt]
  h_2^{\ell\ell} &=&
    \frac{2 \, (1\!-\!u_H)}{(\bar y_H \!-\! u_H)}
   \Big( \big|{\cal C}_{9}\big|^2 \!+\! \big|{\cal C}_{10a}\big|^2
    \!+\! \mbox{Re} \big[{\cal C}_{10a}\, {\cal C}_{10b}^{\,*} \big] \Big)
   \!+\! \frac{\big|{\cal C}_{10b}\big|^2}{2}
  \!-\! \frac{8 \big|{\cal C}_{7}\big|^2 }
  {(1\!-\!\bar y_H)(\bar y_H \!-\! u_H)}  \,,
  \nn \\[5pt]
  h_3^{\ell\ell} &=&
    \frac{-4\,\mbox{Re}[ {\cal C}_{10a}\, {\cal C}_{7}^{\,*}]    }
    {m_B (1-\bar y_H)(\bar y_H - u_H)} -
  \frac{  2\,\mbox{Re}[{\cal C}_{10a}\, {\cal C}_{9}^{\,*}]
    }
    {m_B (\bar y_H - u_H)}
     \,.\nn
\end{eqnarray}
%
%
The full expressions for the coefficients ${\cal C}_{7,9,10a,10b}$
are given in Ref.~\cite{Lee:2005pk}. When we ignore 
${\cal O}(\alpha_s(m_b))$ corrections, they simplify to
\begin{align}
{\cal C}_9 &= C_{9a} \, = \, C_9^{\rm mix} \,, \\
{\cal C}_7 &= C_{7a} \, = \, \frac{\overline m_b(\mu_0)}{m_B} 
                           C_7^{\rm NDR}(\mu_0) \,, \nn \\
{\cal C}_{10a} &= C_{10a} = C_{10} \,, \nn \\
{\cal C}_{10b} &= 0 \,, \nn
\end{align}
where $\mu_0 \sim m_b$ and
\begin{align} \label{C9mix}
 C_9^{\rm mix}(\mu_0) &=  C_9^{\rm NDR}(\mu_0)
 + \frac{2}{9} \left( 3 C_3 + C_4 + 3 C_5 + C_6 \right)
 - \frac{1}{2} h(1, s) \left( 4 C_3 + 4 C_4 + 3 C_5 + C_6 \right)
 \nn \\
&  + h\big(\frac{m_c}{m_b}, s\big)\left( 3 C_1 + C_2 + 3 C_3 + C_4 + 3
C_5 + C_6 \right) - \frac{1}{2} h(0, s) \left( C_3 + 3 C_4 \right)  \nn \\
 & + {\cal O}(\alpha_s(\mu_0)) \,.
\end{align}
The function $h(z,s)$ is given by
\begin{eqnarray}
h(z,s) &=& \frac{8}{9}\ln (\frac{\mu_0}{m_b})
  -\frac{8}{9} \ln z + \frac{8}{27} +\frac{4}{9}\zeta
 -\frac{2}{9}(2 + \zeta) \sqrt{|1-\zeta|} \nn\\
& & \times \bigg[
 \theta(1-\zeta)\bigg(-i\pi+\ln\frac{1+\sqrt{1-\zeta}}{1-\sqrt{1-\zeta}}
   \bigg) +\theta(\zeta-1)\: 2 \arctan \frac{1}{\sqrt{\zeta-1}} \bigg]
 \,, \nn \\
h(0,s) & = & \frac{8}{27}+\frac{8}{9}\ln (\frac{\mu_0}{m_b})
              -\frac{4}{9}\ln s + \frac{4}{9}i\pi \,,
\end{eqnarray}
with $\zeta = 4 z^2/s$ and $s = q^2/m_b^2$.

In the expressions above, $C_{1-6},\, C_{7,9}^{\rm NDR},\, C_{10}$ are the
coefficients of the corresponding operators in the effective Hamiltonian
for $b \to s \ell^+\ell^-$ (for which the NLL calculations were done in
Refs.~\cite{bm95,misiak93}), while $C_9^{\rm mix}$ differs from 
$\tilde C_9^{\rm eff}$ of Ref.~\cite{bm95} by only an ${\cal O}(\alpha_s)$ 
piece.
Note that there is a complication in the perturbative power counting.
Above the scale $m_b$, one usually expands in $\alpha_s$, with
$\alpha_s \log (m_W/m_b) = {\cal O}(1)$. Because of mixing with
${\cal O}_{1,2}$, $C_9 \sim  \log (m_W/m_b) \sim 1/\alpha_s$,
whereas $C_{7,10}  \sim  1$. However, numerically 
$| C_9(m_b) | \sim C_{10}$. This problem is exacerbated by the fact 
that in the shape function region only the rate is calculable, not the 
amplitude.  The solution is to use a ``split matching'' procedure,
which decouples the scale dependence above and below $\mu = m_b$
and thereby allows us to consider the coefficients as ${\cal O}(1)$ numbers 
in the latter region \cite{Lee:2005pk}.

At next-to-leading order, we have
\begin{eqnarray} \label{Wfact00c}
W_i^{(2)f}  &=&
  \frac{h_i^{0f}(\bn\mcdot p)}{2m_b} \:
  \int_{0}^{p_X^+} \!\! dk^+\
  {\cal J}^{(0)}(\bn\mcdot p\, k^+,\mu )\:
   f^{(2)}_0\big(k^+ + r^+,\mu\big) \,
  \nn\\
 &+&
  \sum_{r=1}^2
  \frac{h_i^{rf}(\bn\mcdot p)}{m_b} \:
  \int_{0}^{p_X^+} \!\! dk^+\
  {\cal J}^{(0)}(\bn\mcdot p\, k^+,\mu )\:
   f^{(2)}_r\big(k^+ + r^+,\mu\big) \,
  \nn\\
 &+&
  \sum_{r=3}^4
  \frac{h_i^{rf}(\bn\mcdot p)}{m_b}
  \int\!\! dk_1^+\: dk_2^+
   {\cal J}^{(-2)}(\bn\mcdot p\, k_j^+,\mu )\:
   f^{(4)}_r\big(k_j^+ + r^+,\mu\big) \,
 \nn\\
 &+&
  \sum_{r=5}^6
    \frac{h_i^{rf}(\bn\mcdot p)}{\bn\mcdot p}
  \int\!\! dk_1^+ dk_2^+ dk_3^+  \
 {\cal J}^{(-4)}(\bn\mcdot p\, k_{j'}^+,\mu )\:
   f^{(6)}_r\big(k_{j'}^+ + r^+,\mu\big) 
 \nn\\
 &+& \ldots \,,
\end{eqnarray}
where $r^+ = \bar\Lambda - p_X^+$, $j = 1,2$ and $j' = 1,2,3$.  
The ellipses denote terms that have jet 
functions ${\cal J}$ that start at one-loop order or higher. 
(These terms are given in Ref.~\cite{klis04}.)
When we keep the full dependence on $u_H$, the $h_{1-3}^{ru}$ are
\begin{eqnarray}
  \label{eq:NLOhu13}
h_1^{1u} \!\!& = &\!\! \frac{1}{8} \,, \quad\ \:
h_2^{1u}   =   -\frac{(1-u_H)(2 - \bar y_H - u_H)}{2 (\bar y_H -u_H)^2} 
          \,, \quad\ \
h_3^{1u}   =   \frac{1}{4 m_B (\bar y_H - u_H)} \,, \quad\!\!
   \\[5pt]
h_1^{2u} \!\!& = &\!\! -\frac{1}{4} \,, \quad\!
h_2^{2u}   =   \frac{(1-u_H)((4-u_H) \bar y_H \!-\!\bar y_H^2 \!-\!2)}
              {\bar y_H(\bar y_H-u_H)^2}, \quad\!\!
h_3^{2u}   =   -\frac{1}{2 m_B (\bar y_H - u_H)} , \quad
  \nn \\[5pt]
h_1^{3u} \!\!& = &\!\! -\frac{1}{4 \bar y_H} \,, \quad\!\!
h_2^{3u}   =   -\frac{(1-u_H)}{\bar y_H (\bar y_H - u_H)} \,, 
         \quad\ \ \ \ \ \ \ \
h_3^{3u}   =  -\frac{1}{2 m_B \bar y_H (\bar y_H - u_H)} , \quad 
  \nn \\[5pt]
h_1^{4u} \!\!& = &\!\! \frac{1}{4 \bar y_H} \,, \quad\!\!
h_2^{4u}   =   -\frac{(1-u_H)(2\!-\!\bar y_H\!-\!u_H)}
         {\bar y_H(\bar y_H-u_H)^2} \,, \quad\ \ \ \ \
h_3^{4u}   =   \frac{1}{2 m_B \bar y_H (\bar y_H - u_H)} \,, \quad 
  \nn \\[5pt]
h_1^{5u} \!\!& = &\!\! -\frac{1}{2} \,, \quad
h_2^{5u}   = \frac{2(1-u_H)(1-\bar y_H)}{(\bar y_H - u_H)^2}  \,, \quad\ \ \
h_3^{5u}   =  -\frac{1}{m_B (\bar y_H - u_H)} \,, \quad \: 
  \nn \\[5pt]
h_1^{6u} \!\!& = &\!\! 0 \,, \quad\  \ \
h_2^{6u}   =  \frac{2(1-u_H)^2}{(\bar y_H - u_H)^2} \,, 
      \quad\ \ \ \ \ \ \ \ \ \:
h_3^{6u}   =  0 \,,\quad \ \ \ \ \  \nn 
\end{eqnarray}
and the $h_{1-3}^{r\ell\ell}$ are
\begin{eqnarray}
  \label{eq:NLOhll}
h_1^{1\ell\ell} & = &
 - \, \frac{4|C_{7a}|^2 -
  (|C_{10a}|^2 + |C_{9a}|^2)
  (1-\bar y_H)^2}{4(1-\bar y_H)^2}\
 \, , \\
h_2^{1\ell\ell}   & = &  \frac{(2-\bar y_H-u_H)\left(4|C_{7a}|^2
  - (|C_{10a}|^2 + |C_{9a}|^2)
    (1-\bar y_H)(1-u_H)\right)}
  {(1-\bar y_H)(\bar y_H-u_H)^2} \,, \nn \\
h_3^{1\ell\ell}  & = & - \frac{ \mbox{Re}[C_{10a}\,
                                  C_{9a}^{\,*}]}
  {m_B (\bar y_H - u_H)}   \,, \quad\!\!
  \nn \\[5pt]
h_1^{2\ell\ell} & = &
  \frac{4|C_{7a}|^2 -
  (|C_{10a}|^2 + |C_{9a}|^2)
  (1-\bar y_H)^2}{2(1-\bar y_H)^2}\
  \,, \nn \\
h_2^{2\ell\ell}  & = & -\frac{2}{\bar y_H (\bar y_H - u_H)^2} \bigg[
     4|C_{7a}|^2 \, \frac{2-\bar y_H^2-\bar y_H u_H}{1-\bar y_H}
  + 4 \mbox{Re}[C_{7a}\,C_{9a}^{\,*}]
      (2-\bar y_H-u_H)
 \nn \\
& & \qquad\qquad\qquad
  + \, (|C_{10a}|^2
     + |C_{9a}|^2)(2-4\bar y_H + \bar y_H^2 + \bar y_H u_H)(1-u_H)
    \bigg] \, , \nn \\
h_3^{2\ell\ell}  & = &  \frac{2 \mbox{Re}[C_{10a}\,
                                  C_{9a}^{\,*}]}
  { m_B (\bar y_H - u_H)}    \, ,
  \nn \\[5pt]
h_1^{3\ell\ell} & = & - \, \frac{4|C_{7a}|^2 +
   4 \mbox{Re}[C_{7a}\,C_{9a}^{\,*}]
  (1-\bar y_H)
  + (|C_{10a}|^2 + |C_{9a}|^2)
  (1-\bar y_H)^2}{2\bar y_H(1-\bar y_H)^2}\
\,, \nn \\
h_2^{3\ell\ell}  & = & 2 \, \frac{4|C_{7a}|^2 -
   (|C_{10a}|^2 + |C_{9a}|^2)
   (1-\bar y_H)(1-u_H)}
     {\bar y_H(1-\bar y_H)(\bar y_H - u_H)}
   \,, \nn \\
h_3^{3\ell\ell}  & = & 2\, \frac{2 \mbox{Re}[ C_{10a}\,
                                        C_{7a}^{\,*}]
    +  \mbox{Re}[C_{10a}\,
                 C_{9a}^{\,*}](1-\bar y_H)}
    {m_B \bar y_H (1-\bar y_H) (\bar y_H - u_H)}
  \,,  \nn 
\end{eqnarray}
\begin{eqnarray}
h_1^{4\ell\ell} & = & \frac{4|C_{7a}|^2 +
   4 \mbox{Re}[C_{7a}\,C_{9a}^{\,*}]
  (1-\bar y_H)
  + (|C_{10a}|^2 + |C_{9a}|^2)
  (1-\bar y_H)^2}{2\bar y_H(1-\bar y_H)^2}\
  \,, \nn \\
h_2^{4\ell\ell}  & = & -\frac{2}{\bar y_H (\bar y_H - u_H)^2} \bigg[
     4|C_{7a}|^2 \, \frac{2-\bar y_H - u_H}{1-\bar y_H}
 + 8 \mbox{Re}[C_{7a}\,C_{9a}^{\,*}](1-u_H)
\nn \\
& & \qquad\qquad\qquad
 + \, (|C_{10a}|^2 + |C_{9a}|^2)
   (2-\bar y_H-u_H)(1-u_H) \bigg]
\,, \nn \\
h_3^{4\ell\ell}  & = & -2 \, \frac{2 \mbox{Re}[ C_{10a}\,
                                       C_{7a}^{\,*}]
    + \mbox{Re}[C_{10a}\,
                 C_{9a}^{\,*}](1-\bar y_H)}
    {m_B \bar y_H (1-\bar y_H)(\bar y_H - u_H)}
  \,, \nn \\[5pt]
h_1^{5\ell\ell} & = & - \, \frac{ 4|C_{7a}|^2 +
   4 \mbox{Re}[C_{7a}\,C_{9a}^{\,*}]
  (1-\bar y_H)
  + (|C_{10a}|^2 + |C_{9a}|^2)
  (1-\bar y_H)^2}{(1-\bar y_H)^2}\
 \,, \nn \\
h_2^{5\ell\ell}  & = & 4\,\frac{\left( 4|C_{7a}|^2 +
  4 \mbox{Re}[ C_{7a}\,C_{9a}^{\,*}]
  (1-\bar y_H)
  + (|C_{10a}|^2 + |C_{9a}|^2)
  (1-\bar y_H)^2 \right)(1-u_H)}{(1-\bar y_H)(\bar y_H-u_H)^2}\
 \,, \nn \\
h_3^{5\ell\ell}  & = & 4\, \frac{2 \mbox{Re}[C_{10a}\,
                                    C_{7a}^{\,*}]
    + \mbox{Re}[C_{10a}\,
                 C_{9a}^{\,*}](1-\bar y_H)}
    {m_B (1-\bar y_H)(\bar y_H-u_H)}  \,,
  \nn \\[5pt]
h_1^{6\ell\ell} & = & 0 \,, \nn \\
h_2^{6\ell\ell} & = &  4\,\frac{4|C_{7a}|^2 +
  4 \mbox{Re}[C_{7a}\,C_{9a}^{\,*}] (1-u_H)
  + (|C_{10a}|^2 + |C_{9a}|^2)(1-u_H)^2}
    {(\bar y_H-u_H)^2}\,,
   \nn \\
h_3^{6\ell\ell} & = & 0 \,.\nn
\end{eqnarray}

\section{Weight functions} \label{app:Ku}

%
%
\begin{table}[htb!]
\fbox{
\begin{minipage}{12cm}
{\normalsize
\begin{align}
(4)\qquad K^u_{\rm I\!V} &= \frac{1}{N(u_H)}
\frac{(2x_H+u_H+\bar y_H-2)}{(\bar y_H-u_H)}
\frac{(1-\bar y_H)(2\bar y_H-u_H-1)}{(1+u_H-\bar y_H)}
\nn \\
N(u_H) &= \frac{1}{30}(1-u_H)^2(1-14u_H-94u_H^2-14u_H^3+u_H^4)
- 2 u_H^2(1-u_H^2)\log u_H \nn \\
\kappa_1^u &= \frac{1}{2}\,,\qquad\quad
\kappa_2^u = -1 \nn \\
(5)\qquad K^u_{\rm I\!V} &= 
\frac{6\big[-\!(7-u_H)\bar y_H + 4 x_H \bar y_H + 6 \bar y_H^2 \big]}
{(1-u_H)^7}
\nn \\
\kappa_1^u &= -\frac{1}{10}\,,\qquad
\kappa_2^u = \frac{7-u_H}{5(1-u_H)} \nn \\
(6)\qquad K^u_{\rm I\!V} &= -\frac{105}{101}
\frac{\big[14(1-u_H)^2\bar y_H + 5(2+7u_H) x_H \bar y_H 
      - 45 x_H \bar y_H^2 \big]}{(1-u_H)^8}
\nn \\
\kappa_1^u &= -\frac{2}{101}\,,\qquad
\kappa_2^u = \frac{109-4u_H}{101(1-u_H)} \nn 
\end{align}
}
\end{minipage}
}
\caption{Some choices of $K^u(x_H,\bar y_H,u_H)$ for which 
the weighted integral Eq.~(\ref{eq:Ku}) depends only on the
shape functions $F$, $F_1$ and $F_2$.
The coefficients $\kappa_1^u(u_H)$ and $\kappa_2^u(u_H)$ 
are defined in Eq.~(\ref{eq:Musll}).
}
   \label{table:012}
\end{table}
\begin{table}[htb!]
\fbox{
\begin{minipage}{12cm}
{\normalsize
\begin{align}
(7)\qquad K^{\ell\ell} &= \frac{1}{N(u_H)}
\frac{(2x_H+u_H+\bar y_H-2)}{(\bar y_H-u_H)}
\frac{(\bar y_{H*}\!-\!\bar y_H\!-\!u_H)^2}{(\bar y_{H*}-\bar y_H)}
\nn \\
&\qquad \times
\left\{
{\cal A} - {\cal B}[1-(\bar y_{H*}\!-\!\bar y_H)^2]
\right\}
(\bar y_{H*}-2\bar y_H)
\nn \\[2pt]
N(u_H) &= 8(1-u_H) \int_{\bar y_{\rm min}}^{\bar y_{\rm max}} d\bar y_H 
(\bar y_H\!-\!u_H)^2 
\frac{(\bar y_{H*}\!-\!\bar y_H\!-\!u_H)^2}{(\bar y_{H*}-\bar y_H)}
(\bar y_{H*}-2\bar y_H)
\nn \\
&\qquad\qquad\qquad\qquad \times
\left\{
{\cal A} - {\cal B}(1-\bar y_H)
\right\}
\left\{
{\cal A} - {\cal B}[1-(\bar y_{H*}\!-\!\bar y_H)^2]
\right\}
\nn \\[2pt]
\kappa_2^{\ell\ell} &= -\frac{8(1-u_H)}{N(u_H)} 
\int_{\bar y_{\rm min}}^{\bar y_{\rm max}} d\bar y_H 
(\bar y_H\!-\!u_H)^2 (1-\bar y_H)
\frac{(\bar y_{H*}\!-\!\bar y_H\!-\!u_H)^2}{(\bar y_{H*}-\bar y_H)}
(\bar y_{H*}-2\bar y_H)
\nn \\
&\qquad\qquad\qquad\qquad \times
   {\cal B} 
\left\{
{\cal A} - {\cal B}[1-(\bar y_{H*}\!-\!\bar y_H)^2]
\right\}
\nn \\[6pt]
(8)\qquad K^{\ell\ell} &= \frac{1}{N(u_H)}
(2x_H+u_H+\bar y_H-2)
(\bar y_{H*}\!-\!\bar y_H\!-\!u_H)^3\bar y_H
(\bar y_{H*}-2\bar y_H)
\nn \\
&\qquad\qquad\qquad\qquad \times
\left\{
{\cal A} - {\cal B}[1-(\bar y_{H*}\!-\!\bar y_H)^2]
\right\}
\nn \\[2pt]
N(u_H) &= 8(1-u_H) \int_{\bar y_{\rm min}}^{\bar y_{\rm max}} d\bar y_H
(\bar y_H\!-\!u_H)^3
(\bar y_{H*}\!-\!\bar y_H\!-\!u_H)^3 \bar y_H
(\bar y_{H*}-2\bar y_H)
\nn \\
&\qquad\qquad\qquad\qquad \times
\left\{
{\cal A} - {\cal B}[1-(\bar y_{H*}\!-\!\bar y_H)^2]
\right\}
\left\{
{\cal A} - {\cal B}(1-\bar y_H)
\right\}
\nn \\[2pt]
\kappa_2^{\ell\ell} &= -\frac{8(1-u_H)}{N(u_H)} 
\int_{\bar y_{\rm min}}^{\bar y_{\rm max}} d\bar y_H
(\bar y_H\!-\!u_H)^3 (1-\bar y_H)
(\bar y_{H*}\!-\!\bar y_H\!-\!u_H)^3
\bar y_H(\bar y_{H*}-2\bar y_H)
\nn \\
&\qquad\qquad\qquad\qquad \times
{\cal B}
\left\{
{\cal A} - {\cal B}[1-(\bar y_{H*}\!-\!\bar y_H)^2]
\right\}
\nn \\[6pt]
(9)\qquad K^{\ell\ell} &= \frac{1}{N(u_H)}
\frac{(2x_H+u_H+\bar y_H-2)}{(\bar y_H-u_H)}
\frac{(\bar y_{H*}\!-\!\bar y_H\!-\!u_H)^2}{(\bar y_{H*}-\bar y_H)}
\frac{(\bar y_{H*}-2\bar y_H)}{
\left\{
{\cal A} - {\cal B}(1-\bar y_H^2)
\right\}
}
\nn \\[2pt]
N(u_H) &= 8(1-u_H) \int_{\bar y_{\rm min}}^{\bar y_{\rm max}} d\bar y_H
(\bar y_H\!-\!u_H)^2
\frac{(\bar y_{H*}\!-\!\bar y_H\!-\!u_H)^2}{(\bar y_{H*}-\bar y_H)}
(\bar y_{H*}-2\bar y_H)
\nn \\
&\qquad\qquad\qquad\qquad \times
\frac{\left\{
{\cal A} - {\cal B}(1-\bar y_H)
\right\}}
{\left\{
{\cal A} - {\cal B}(1-\bar y_H^2)
\right\}}
\nn \\[2pt]
\kappa_2^{\ell\ell} &= -\frac{8(1-u_H)}{N(u_H)} 
\int_{\bar y_{\rm min}}^{\bar y_{\rm max}} d\bar y_H
(\bar y_H\!-\!u_H)^2 (1-\bar y_H)
\frac{(\bar y_{H*}\!-\!\bar y_H\!-\!u_H)^2}{(\bar y_{H*}-\bar y_H)}
(\bar y_{H*}-2\bar y_H)
\nn \\
&\qquad\qquad\qquad\qquad \times
\frac{\cal B}
{\left\{
{\cal A} - {\cal B}(1-\bar y_H^2)
\right\}}
\nn 
\end{align}
}
\end{minipage}
}
\caption{Some choices of $K^{\ell\ell}(x_H,\bar y_H,u_H)$ for which
$\kappa_2^{\ell\ell}(u_H) = \kappa_3^{\ell\ell}(u_H)$ in 
Eq.~(\ref{eq:Musll}).\protect\footnotemark[1] 
Here, ${\cal A} = -2\mbox{Re}[ C_{10a}\, C_{7a}^{\,*}]$, 
${\cal B} = \mbox{Re}[C_{10a}\, C_{9a}^{\,*}]$ and 
$\bar y_{H*} = \bar y_{\rm min} + \bar y_{\rm max}$.
}
\footnotetext[1]{Note that Example $(9)$ requires a harsher cut,
e.g.\ $2 \,\mbox{GeV}^2 \leq q^2 \leq 6 \,\mbox{GeV}^2$
(rather than $1 \,\mbox{GeV}^2 \leq q^2 \leq 6 \,\mbox{GeV}^2$), 
so that it is not singular.}
   \label{table:ll23}
\end{table}
%
%
%
\begin{table}[htb!]
\fbox{
\begin{minipage}{12cm}
{\normalsize
\begin{align}
(10)\qquad K^{\ell\ell} &= \frac{1}{N(u_H)}
\frac{(2x_H+u_H+\bar y_H-2)}{4(1-u_H)}
\nn \\[2pt]
N(u_H) &= 2 \int_{\bar y_{\rm min}}^{\bar y_{\rm max}} d\bar y_H
(\bar y_H\!-\!u_H)^3
\left\{
{\cal A} - {\cal B}(1-\bar y_H)
\right\}
\nn \\[2pt]
\kappa_2^{\ell\ell} &= -\frac{2}{N(u_H)} 
\int_{\bar y_{\rm min}}^{\bar y_{\rm max}} d\bar y_H
     {\cal B}
(1-\bar y_H)(\bar y_H\!-\!u_H)^3
\nn \\[2pt]
\kappa_3^{\ell\ell} &= -\frac{2}{N(u_H)} 
\int_{\bar y_{\rm min}}^{\bar y_{\rm max}} d\bar y_H
\frac{(\bar y_H\!-\!u_H)^3}{\bar y_H}
\left\{
{\cal A} - {\cal B}(1-\bar y_H)
\right\}
\nn \\[6pt]
(11)\qquad K^{\ell\ell} &= \frac{1}{N(u_H)}
\frac{(2x_H+u_H+\bar y_H-2)\bar y_H}{4(1-u_H)}
\nn \\[2pt]
N(u_H) &= 2 \int_{\bar y_{\rm min}}^{\bar y_{\rm max}} d\bar y_H \,
\bar y_H(\bar y_H\!-\!u_H)^3
\left\{
{\cal A} - {\cal B}(1-\bar y_H)
\right\}
\nn \\[2pt]
\kappa_2^{\ell\ell} &= -\frac{2}{N(u_H)} 
\int_{\bar y_{\rm min}}^{\bar y_{\rm max}} d\bar y_H
    {\cal B}
(1-\bar y_H)(\bar y_H\!-\!u_H)^3 \bar y_H
\nn \\[2pt]
\kappa_3^{\ell\ell} &= -\frac{2}{N(u_H)} 
\int_{\bar y_{\rm min}}^{\bar y_{\rm max}} d\bar y_H
(\bar y_H\!-\!u_H)^3
\left\{
{\cal A} - {\cal B}(1-\bar y_H)
\right\}
\nn \\[6pt]
(12)\qquad K^{\ell\ell} &= \frac{1}{N(u_H)}
\frac{(2x_H+u_H+\bar y_H-2)}{4(1-u_H)(\bar y_H - u_H)}
\frac{(\bar y_{H*}\!-\!\bar y_H\!-\!u_H)^2}
{(\bar y_{H*}-\bar y_H)}
(\bar y_{H*}-2\bar y_H)
\nn \\
&\qquad\qquad\qquad\qquad \times
\left\{
{\cal A} - {\cal B}(1\!-\!\bar y_{H*}\!+\!\bar y_H)
\right\}
\nn \\[2pt]
N(u_H) &= 2 \int_{\bar y_{\rm min}}^{\bar y_{\rm max}} d\bar y_H
(\bar y_H\!-\!u_H)^2 
\frac{(\bar y_{H*}\!-\!\bar y_H\!-\!u_H)^2}
{(\bar y_{H*}-\bar y_H)} 
(\bar y_{H*}-2\bar y_H)
\nn \\
&\qquad\qquad\qquad\qquad \times
\left\{
{\cal A} - {\cal B}(1-\bar y_H)
\right\}
\left\{
{\cal A} - {\cal B}(1\!-\!\bar y_{H*}\!+\!\bar y_H)
\right\}
\nn \\[2pt]
\kappa_2^{\ell\ell} &= -\frac{2}{N(u_H)} 
\int_{\bar y_{\rm min}}^{\bar y_{\rm max}} d\bar y_H
(1-\bar y_H)(\bar y_H\!-\!u_H)^2 
\frac{(\bar y_{H*}\!-\!\bar y_H\!-\!u_H)^2}
{(\bar y_{H*}-\bar y_H)}
\nn \\
&\qquad\qquad\qquad\qquad \times
(\bar y_{H*}-2\bar y_H)
{\cal B}\left\{
{\cal A} - {\cal B}(1\!-\!\bar y_{H*}\!+\!\bar y_H)
\right\}
\nn \\[2pt]
\kappa_3^{\ell\ell} &= 0 
\nn 
\end{align}
}
\end{minipage}
}
\caption{Some choices of $K^{\ell\ell}$ and the
corresponding coefficients $\kappa_2^{\ell\ell}(u_H)$
and $\kappa_3^{\ell\ell}(u_H)$, which are defined in
Eq.~(\ref{eq:Musll}).
Here, ${\cal A} = -2\mbox{Re}[ C_{10a}\, C_{7a}^{\,*}]$,  
${\cal B} = \mbox{Re}[C_{10a}\, C_{9a}^{\,*}]$ and
$\bar y_{H*} = \bar y_{\rm min} + \bar y_{\rm max}$.
$C_{9a}$ may be taken to be constant, in which case the integrals
can be evaluated analytically.}
   \label{table:ll}
\end{table}
%
%
%
\newpage
\phantom{x}
\newpage
\phantom{x}
\newpage



\end{document}